\documentclass[twocolumn,pra,showpacs,nofootinbib]{revtex4-2}
\usepackage{epsfig}
\usepackage{dcolumn}
\usepackage{natbib}
\usepackage[utf8]{inputenc}
\usepackage[english]{babel}
\usepackage{xcolor}
\usepackage{subcaption}
\usepackage{graphicx}
\usepackage{xcolor}
\usepackage{color}
\usepackage{bm}
\usepackage{tabularx}
\usepackage{amssymb}
\usepackage[intlimits]{amsmath}
\usepackage{booktabs}
\usepackage{amsmath}
\usepackage[para,online,flushleft]{threeparttable}

\definecolor{darkcandyapplered}{rgb}{0.64, 0.0, 0.0}
\usepackage[colorlinks=true,allcolors=darkcandyapplered]{hyperref}
\usepackage{physics}
\usepackage{multirow}
\usepackage{float}
\begin{document}
\title{Constraints on the Variation of the QCD Interaction Scale $\Lambda_{\text{QCD}}$}

\author{V. V. Flambaum} 
\email{v.flambaum@unsw.edu.au}
\author{A. J. Mansour}
\email{andrew.mansour@student.unsw.edu.au}
\affiliation{School of Physics, University of New South Wales,
Sydney 2052, Australia}
\date{\today}

\begin{abstract}

Laboratory and astrophysical tests of ``constant variation'' have so far concentrated on the \textit{dimensionless} fine-structure constant $\alpha$ and on the electron or quark mass ratios $X_{e,q}=m_{e,q}/\Lambda_{\text{QCD}}$, treating the QCD scale $\Lambda_{\text{QCD}}$ as unchangeable. Certain beyond Standard Model frameworks, most notably those with a dark matter or dark energy scalar field $\phi$ coupling with the gluon field, would make $\Lambda_{\text{QCD}}$ itself time dependent while leaving $\alpha$ and the electron  mass untouched. Under the minimal assumption that this gluonic channel is the sole $\phi$ interaction, we recast state-of-the-art atomic clock comparisons into $\dot{\Lambda}_{\text{QCD}}/\Lambda_{\text{QCD}}=(3.2 \pm 3.5) \times 10^{-17} \ \text{yr}^{-1}$ limits, translate the isotope yields of the 1.8-Gyr-old Oklo natural reactor into a complementary geophysical limit of $|\delta\Lambda_{\text{QCD}}/\Lambda_{\text{QCD}}|<2\times10^{-9}$ over that time span, corresponding to the linear drift limit $|\dot{\Lambda}_{\text{QCD}}/\Lambda_{\text{QCD}}|<1\times10^{-18} \text{yr}^{-1}$, and show that the proposed $8.4$ eV $^{229}$Th nuclear clock would amplify a putative $\Lambda_{\text{QCD}}$ drift by four orders of magnitude compared with present atomic clocks. We also obtain constraints from quasar absorption spectra and Big Bang Nucleosynthesis data.

\end{abstract}

\maketitle

\section{Introduction}
The search for the variation of fundamental constants is a well-established area of research motivated by theories beyond the Standard model that unify gravity with the other fundamental forces, as well as cosmological models involving scalar fields which predict a space-time variation of these constants. Constraints on such variation have been derived from Big Bang Nucleosynthesis (BBN), quasar absorption spectra, the cosmic microwave background (CMB) and the Oklo natural nuclear reactor, see e.g. the review~\cite{Uzan} and references therein. The stringent bounds on the linear drift of the fundamental constants have been obtained using measurements of the temporal drift of atomic clock transition frequencies. These experiments measure the variation of the dimensionless ratio of two transition frequencies $\nu_{1}$ and $\nu_{2}$
\begin{equation} \label{ratio}
\delta \left( \frac{\nu_1}{\nu_2} \right) \,,
\end{equation} 
which have different dependencies on the fundamental constants. As a result, these measurements may be used to place constraints on the variation of these constants. Experimental results have been obtained by measuring the oscillating frequency ratios of electron transitions in a range of systems, including Dy/Cs \cite{DyCs}, Rb/Cs \cite{Hees2016}, Yb/Cs \cite{YbCs}, Sr/H/Si cavity \cite{HSi}, Cs/cavity \cite{Tretiak}, Cs/H~\cite{Fischer2004}, Al$^{+}$/Hg$^{+}$~\cite{AlHgdrift} and Yb$^+$/Yb$^+$/Sr \cite{Banerjee2023,Filzinger2023}. The results of these experiments have been used to place constraints on the variations of fundamental constants using the sensitivity calculations performed in Refs.~\cite{PRLWebb,PRAWebb,CanJPh,Thomas,Tedesco,csquarks,Borschevsky} and several other papers referenced there.

It is important to note that if the absolute measurement of a transition frequency is performed, the dimensionless ratio of this frequency to the Cs hyperfine transition frequency, which defines the units of frequency and time, is being measured. The Cs transition frequency has a complicated dependence on the fundamental constants~\cite{Tedesco}. This principle applies to any measurement of variations of fundamental constants in any field of physics and astrophysics. Any measurement of a dimensionful parameter is, in fact, a measurement of the dimensionless ratio of that parameter to the unit used to measure it, which itself depends on the fundamental constants.   

Because any measurable quantity is ultimately a dimensionless ratio, it can always be written solely in terms of dimensionless fundamental constants. 
To do so, one first identifies the particular dimensionless constants that enter a given experiment. 
By contrast, many papers discuss variations of dimensionful parameters, such as the speed of light $c$, Planck’s constant $\hbar$, Newton’s gravitational constant $G_N$, the QCD scale $\Lambda_{\text{QCD}}$, or particle masses. 
The problem is that the numerical values of these quantities depend on the system of units: in natural units, for example, $c=1$ by definition, so it cannot "vary."  Any apparent change can be absorbed by a re-definition of units.
Consequently, interpreting experimental results in terms of changing dimensionful constants is not model-independent; one must instead identify a set of dimensionless fundamental constants relevant to this measurement and calculate the dependence of an observed effect on these constants \footnote{It was noted in Refs. \cite{Antypas1,Antypas2}, that the number of possible "dimensionless" constants may increase when we allow for fast variations of the constants, where "fast" is determined by the time scale of the response of the studied species or experimental apparatus used. In this case, the relevant dimensionless quantity is, for example, the ratio $m_e/\ev{m_e}$, and $\ev{m_e}$ is the time average. In this sense, one may say that the experimental signal depends on the variation of dimensionful constants (the electron mass $m_e$ in this example). Sensitivity to
such variations requires two systems, one which has a
faster response (such as an atom) and another which is more
inertial (such as a cavity).}



That said, fundamental Lagrangians are often written with dimensionful constants. 
To be consistent, one must include every relevant term in the Lagrangian when deriving how the dimensionless ratios of fundamental constants vary for a given measurement.



Alternatively, one may speak about the apparent variation of the fundamental constants, which may be just a convenient way to calculate the effects of specific terms in the interaction of Standard model particles with a dark matter (or dark energy \cite{BarrowMagueijo1998,Barrow2005,BarrowSandvikMagueijo2002,Wetterich2003,Uzan}) scalar or pseudoscalar field $\phi$. For example, the interaction of fermions with $\phi$, which may be presented in the form $g \phi^n m \bar f f $, is similar to the fermion mass term $ m \bar f f $. This means that one may define a variable fermion effective mass $m_{eff} = m(1+ g \phi^n)$, which depends on the field $\phi$. Here, $n=1,2$ for scalars and $n=2$ for pseudoscalars. Similarly, the interaction of $\phi$ with the electromagnetic field may be incorporated as a variable fine structure constant $\alpha_{eff}=\alpha  (1+ g \phi^n)$. The effects of the variation of fermion masses and the variation of $\alpha$ have been calculated; see, e.g. 
\cite{PRLWebb,PRAWebb,CanJPh,Shuryak2002,FlambaumShuryak,Thomas,Tedesco,csquarks,Borschevsky}. Note that in this approach, the variation of measurement units is still a relevant issue if one considers measurements of dimensionful parameters. We note that using experimental results to place constraints on the variation of dimensionful parameters is only possible within a specific model of variation, e.g. an interaction with an ultralight dark matter field or dark energy field, which varies on a cosmological scale.

For clarity, we note that probing an extremely small fractional change in a constant does not require knowing that constant’s absolute value with comparable precision. What matters is the sensitivity coefficient that links observable shifts to fractional changes. For example, if an atomic transition frequency $\nu$ obeys $\delta \nu/\nu= k_{\alpha} (\delta \alpha/\alpha)$
one only needs a  calculation of  $k_{\alpha}$ and high precision measurement of $\delta \nu/\nu$. At present  $\delta \alpha/\alpha$ is constrained to better than $10^{-18}$ per year, even though the absolute value of $\alpha$  is known only to $1.5 \times 10^{-10}$.

The model-independent variations of dimensionless constants can be attributed to the fractional variation of three fundamental independent dimensionless parameters: the fine structure constant $\alpha$ and the ratios of the electron/quark masses to the QCD scale $\Lambda_{\text{QCD}}$, $X_{e}$ and $X_{q}$, where $\Lambda_{\text{QCD}}$ is defined as the position of the Landau pole in the logarithm for the running strong coupling constant, $\alpha_{s} (r) \sim \text{const}/ \ln (\Lambda_{\text{QCD}} r / \hbar c)$.

We seek to examine the sensitivity of various systems to variations of the QCD scale $\Lambda_{\text{QCD}}$. Such a sensitivity appears in the frequency comparison of two microwave atomic clocks or optical and microwave  clocks (see below). 
It is usually assumed that the comparison of two optical clock transitions is sensitive to variations of the fine structure constant only; however,
following Ref. \cite{Banerjee2023},
we argue that this system is also sensitive to variations of $\Lambda_{\text{QCD}}$ and repurpose existing results to place constraints. We also examine the sensitivity of the fractional variation of the transition frequency between the ground state and the low-energy isomer in $^{229}$Th to variations in the QCD scale $\Lambda_{\text{QCD}}$, which appears to be enhanced relative to the proposed nuclear clock's sensitivity to the quark mass $m_{q}$. Additionally, we place constraints on $\Lambda_{\text{QCD}}$ from quasar absorption spectra and Big Bang Nucleosynthesis. Finally, we reinterpret existing constraints from the Oklo natural nuclear reactor to place constraints on the variation of $\Lambda_{\text{QCD}}$.


\section{Constraints on $\Lambda_{\text{QCD}}$ from  atomic clocks measurements}

\subsection{Ratios of magnetic dipole hyperfine  and optical transition  frequencies}

The experimental observation of variation in nature may be interpreted in a model-independent manner as the variation of at least one of three independent fundamental dimensionless ratios;

\begin{align} \label{dimensionlessratios}
    \frac{\delta \alpha}{\alpha}\,, \ \frac{\delta X_{q}}{X_{q}}\,, 
 \ \text{and} \ \frac{\delta X_{e}}{X_{e}}\,,
\end{align}
where $X_{q,e}$ corresponds to the ratio of the quark/electron mass to the QCD scale $\Lambda_{\text{QCD}}$.
Here, we are working in natural units, $c = \hbar = 1$.
For example, the variation of the ratio of two atomic, molecular or nuclear transition frequencies may be presented as
\begin{align}
    \frac{\delta (\nu_1/\nu_2)}{\nu_1/\nu_2} = k_{\alpha} \frac{\delta \alpha}{\alpha} + k_{q} \frac{\delta X_{q}}{X_{q}} + k_{e} \frac{\delta X_{e}}{X_{e}} \,,
\end{align}
Table \ref{Clockvariations} lists the current constraints on the variation of the ratio of the two hyperfine transition frequencies in $^{87}$Rb and $^{133}$Cs and the constraints on the variation in the ratio of optical transition frequencies in $^{1}$H, $^{199}$Hg$^{+}$, $^{171}$Yb$^{+}$ and $^{87}$Sr to the hyperfine transition in $^{133}$Cs. In the previous calculations in Refs.~\cite{Thomas,Tedesco}, the dependence on $X_q$ was calculated assuming that $\Lambda_{\text{QCD}}$ is constant and only the quark masses vary. Below, we will demonstrate that there is no need for such an assumption.

The dependence of the atomic hyperfine transition frequency on $X_q$ appears mainly  from the dependence on the nuclear magnetic moment. The dependence of the nuclear magnetic moments on $m_q$ was calculated in Refs.~\cite{Thomas}.  A less significant contribution to the $m_q$ dependence, which originates from the variation of the nuclear radius, was calculated in Refs. \cite{Wiringa2009,Dinh2009}.

The nuclear magnetic moment and nuclear radius are also sensitive to the QCD scale $\Lambda_{\text{QCD}}$. In effective field theory, the nucleon magnetic moments receive loop-level corrections from pion-nucleon interactions governed by the spontaneously broken chiral symmetry of QCD. These corrections provide a natural mechanism by which these magnetic moments can acquire a dependence on both the quark mass and $\Lambda_{\text{QCD}}$. The nucleon magnetic moment including such corrections can be schematically written as~\cite{Bernard1995,Kubis2001}:
\begin{equation}
    \mu_N = \mu_N^{(0)} + \mu_{N,\pi}^{(1)} \,,
\end{equation}
where $\mu_N^{(0)}$ denotes the static low-energy constant contributions at tree-level and $\mu_{N,\pi}^{(1)}$ represents the dominant corrections to the nucleon magnetic moments, measured in units of the nuclear magneton $e\hbar/2 m_p c$,  which at leading one-loop order are proportional to the pion mass $m_{\pi}$~\cite{Bernard1992,Jenkins1993}:
\begin{equation}\label{muN}
    \mu_{N,\pi}^{(1)} \sim \frac{g_A^2 m_\pi m_N}{8\pi f_\pi^2} \,,
\end{equation}
where $g_A$ is the axial coupling, $m_{N}$ the nucleon mass, and $f_\pi$ the pion decay constant. The sign of this correction is positive for the proton magnetic moment and negative for the neutron magnetic moment. Since $m_\pi^2 \sim m_q \Lambda_{\text{QCD}}$  via the Gell-Mann--Oakes--Renner relation ~\cite{GellMan1968} and $f_{\pi} \propto \Lambda_{\text{QCD}}$, the pion mass inherits a dependence on the light quark mass $m_q$ and $\Lambda_{\text{QCD}}$. Consequently, the one-loop correction scales as:
\begin{equation}\label{muN1}
    \mu_{N,\pi}^{(1)} \propto \left( \frac{m_q}{\Lambda_{\text{QCD}}} \right)^{1/2}\,.
\end{equation}
Thus, we see that nucleon magnetic moments are naturally sensitive probes of variations of dimensionless ratio of  the quark mass $m_{q}$ and $\Lambda_{\text{QCD}}$.
\begin{table}[h]
    \centering
    \resizebox{\columnwidth}{!}{%
    \begin{tabular}{ccccc}
    \hline
    \hline
    Clock 1 & Clock 2 & Constraint (yr$^{-1}$) & Ref. & Constants dependence  \\
    \hline
      $^{87}$Rb  & $^{133}$ Cs & $(-1.4 \pm 0.9) \times 10^{-16}$  & \cite{RbCsNew} & $\alpha^{-0.49} X_{q}^{-0.025}$ \\
      
      $^{1}$H & $^{133}$Cs & $(-32 \pm 63) \times 10^{-16}$ &  \cite{FischerHCs2004} & $\alpha^{2.83} X_{q}^{-0.039} X_{e} $ \\
      
      $^{199}$Hg$^{+}$ & $^{133}$Cs & $(3.7 \pm 3.9) \times 10^{-16}$ & \cite{Fortier2007} & $\alpha^{6.05} X_{q}^{-0.039} X_{e}$ \\

      $^{171}$Yb$^{+}$ & $^{133}$Cs & $(-3.1 \pm 3.4) \times 10^{-17}$ & \cite{Lange2021} & $\alpha^{1.93} X_{q}^{-0.039} X_{e}$ \\

      $^{87}$Sr & $^{133}$Cs & $(-1.0 \pm 1.8) \times 10^{-15}$ & \cite{Blatt2008} & $\alpha^{2.77} X_{q}^{-0.039} X_{e}$ \\

      \hline
      \hline
    \end{tabular}
    }
    \caption{Constraints on the variation of the ratio of clock frequencies.}
    \label{Clockvariations}
\end{table}


Certain beyond-Standard Model frameworks - most notably those with an ultralight scalar field $\phi$ - permit a direct scalar-gluon coupling of the form $\phi^n G_{\mu\nu}^aG^{a\mu\nu}$. Such an interaction would make $\Lambda_{\text{QCD}}$ time-dependent while leaving $\alpha$ and electron mass untouched (neglecting radiative corrections). If $\phi$ is a dark energy field, which evolves on cosmological timescales, this coupling leads to a slow drift of $\Lambda_{\text{QCD}}$ for $n=1$ or $n=2$. If instead $\phi$ corresponds to an ultralight dark matter field, it oscillates as $\phi_0\cos{(mt)}$, and a gradual variation of $\Lambda_{\text{QCD}}$ may arise for $n=2$ through the evolution of the average value of the dark matter density $\propto \phi^2$.

Under the minimal assumption that this gluonic channel is the sole $\phi$ interaction, 
the constraints presented in Table~\ref{Clockvariations} can be translated into limits on the temporal variation of the QCD scale, $\Lambda_{\text{QCD}}$, via the dependence of each transition on $X_{q}$ and $X_{e}$.
The dominant contribution to the sensitivity arises from the $X_{e}$ term, and the tightest constraints may be obtained from the $^{171}$Yb$^{+}$/Cs clock measurement~\cite{Lange2021}:

\begin{align}
    \frac{\delta \Lambda_{\text{QCD}}}{\Lambda_{\text{QCD}}} = (3.2 \pm 3.5) \times 10^{-17} \ \text{yr}^{-1} \,.
\end{align}


\subsection{Ratio of electric quadrupole hyperfine to  magnetic dipole hyperfine  and optical transition  frequencies}  

In our paper \cite{FlambaumMansour2023} we found that the  electric quadrupole hyperfine constant $B$ has higher sensitivity to variation of $m_q$ than the magnetic hyperfine constant  $A$, and suggested to measure variation of the ratio $B/A$. In this subsection we will show that  this ratio $B/A$, as well as the ratio of $B$ to an optical transition frequency, are highly sensitive to the variation of $\Lambda_{\text{QCD}}$.  

The dependence of the  hyperfine constants  $B$ and $A$ on the fundamental constants is given by:
\begin{align} 
\label{B}
B & \propto  \frac{e Q}{a_B^3} R_B(Z\alpha) \sim \frac{e^2 r_0^2 }{a_{B}^3}R_B(Z\alpha) \,,\\
\label{A}
A & \propto  \frac{\mu_B \mu_N}{a_{B}^3}R_A(Z \alpha) \sim \frac{ e^2 \hbar^2 g_N}{m_e m_N c^2 a_{B}^3} R_A(Z \alpha) \,,
\end{align}
where $r_{0} \approx 1.2 \ \text{fm}$ is the internucleon distance, $a_{B}$ is the Bohr radius, $\mu_B$ and  $\mu_N$ are the Bohr and nuclear magnetons and $g_N$ is the nuclear magnetic $g$-factor, $Q$ is the electric quadrupole moment, $R_B(Z\alpha)$  and  $R_A(Z \alpha)$ are the relativistic factors for the electric quadrupole and magnetic dipole constants respectively, which are presented e.g. in the paper~\cite{Tedesco} and the book~\cite{Sobelman}.  Therefore, the ratio of the electric quadrupole and magnetic dipole constants may be written as  
\begin{align} \label{BA}
\frac{B}{A} \propto  \frac{ r_0^2 m_e m_N c^2 R_B(Z\alpha)}{g_N \hbar^2 R_A(Z \alpha)} \,. 
\end{align}
The dominating effect in the variation of this ratio comes from the variation of $r_{0}^{2}$ and $m_N$. The dependence of $g_N$ on hadronic parameters is generally weak and varies from nucleus to nucleus~\cite{Thomas,Tedesco}. To leading order, this dependence can be neglected, allowing for general estimates that are valid across all nuclei. 

The variation of the internucleon distance $r_0$ in terms of the pion and nucleon mass has been calculated in Ref.~\cite{Wiringa2009}
\footnote{The dependence on the nucleon and pion masses arises mainly from the kinetic energy and the long-distance behavior of the strong interaction. These contributions can be calculated reliably and are included in the present work. Ref.~\cite{Wiringa2009} also considered vector meson and $\Delta$ contributions, which describe short-distance nucleon interaction. The description of the short- distance interaction is model-dependent. Also, the $V$ and $\Delta$ contributions partially cancel. Since their inclusion does not clearly improve the accuracy, we omit the $V$ and $\Delta$ contributions.}:
\begin{align} \label{rpi}
 \frac{\delta r_0}{r_0}=1.8  \frac{\delta m_{\pi}}{m_{\pi}} -  4.8 \frac{\delta m_N}{m_N}\,.
\end{align} 
The variation of the nucleon mass $m_{N}$ can be related to the variation of the fundamental constants by~\cite{Thomas,Damour2010,SHIFMAN1978,Junnarkar2013,Aoki2020,Oswald2022}

\begin{align} \label{LambdaQCDFC}
\begin{split}
    \frac{\delta m_{N}}{m_{N}} & = 0.909 \frac{\delta \Lambda_{\text{QCD}}}{\Lambda_{\text{QCD}}} + 0.084\frac{\delta m_{q}}{m_{q}} \\
    & + 3 \times 10^{-4} \frac{\delta m_{q,-}}{m_{q,-}} + 0.043 \frac{\delta m_{s}}{m_{s}} \,,
\end{split}
\end{align}
where $m_{q,-} = (m_{u} - m_{d})$ is the mass difference.
The dependence of the variation of the nucleon mass on this parameter is weak, and we may neglect it. We will also assume that $\frac{\delta m_{s}}{m_{s}}=\frac{\delta m_{q}}{m_{q}}$, where $m_{s}$ is the mass of the strange quark (this assumption may be motivated by the Higgs mechanism of mass generation). Noting the dependence of the pion mass on $\Lambda_{\text{QCD}}$ and $m_{q}$ from the Gell-Mann-Oakes-Renner relation~\cite{GellMan1968}

\begin{align} \label{GMOR}
    m_{\pi}^{2} \sim \Lambda_{\text{QCD}} m_{q} \,,
\end{align}
we yield the following dependence on the hadronic parameters

\begin{align} \label{deltar0Lambda}
    \frac{\delta r_{0}}{r_{0}} = -3.5 \frac{\delta \Lambda_{\text{QCD}}}{\Lambda_{\text{QCD}}} + 0.3 \frac{\delta m_{q}}{m_{q}} \,.
\end{align}
Thus, using Eqs. (\ref{BA}) and (\ref{deltar0Lambda}), we yield

\begin{align}
    \frac{\delta(B/A)}{(B/A)} = -6.0 \frac{\delta \Lambda_{\text{QCD}}}{\Lambda_{\text{QCD}}} + 0.7 \frac{\delta m_{q}}{m_{q}} \,.
\end{align}
It may also be useful to estimate the dependence of the ratio of the electric quadrupole constant $B$ to the optical transition energy $E_\text{opt} \propto e^2/a_B$. Using Eq. (\ref{B}) we obtain
\begin{align} \label{Bopt}
 \frac{\delta(B/E_\text{opt})}{(B/E_{opt})}  \approx \frac{\delta (r_{0}^2/a_B^2)}{r_{0}^2/a_B^2}.
\end{align}
We omit here the relativistic factors which contribute to the $\alpha$-dependence. The dependence on hadronic parameters may be presented as 
\begin{align} \label{BoptH}
 \frac{\delta(B/E_\text{opt})}{(B/E_{opt})}=
  -9.6 \frac{\delta m_N}{m_N} + 3.6 \frac{\delta m_{\pi}}{m_{\pi}} \\
  = -6.9 \frac{\delta \Lambda_{\text{QCD}}}{\Lambda_{\text{QCD}}} + 0.6 \frac{\delta m_{q}}{m_{q}}.
\end{align}

\subsection{Sensitivity of the comparison of two optical clock transition frequencies to hadronic parameters}

Different types of atomic transition frequencies have a differing dependence on the fundamental constants. As discussed above, constraints obtained from the measurement of the ratio of optical and hyperfine transition frequencies have sensitivity to $X_{e}$ via a dependence on the electron to proton mass ratio $m_{e}/m_{p}$ and sensitivity to $X_{q}$ via dependence on the nuclear magnetic g-factor. 

Following Ref. \cite{Banerjee2023}, we aim to propose a similar dependence on $X_{e}$ for the ratio of two optical frequencies, by probing the contributions of nuclear mass $m_{N}$  and nuclear radius $r_{N}$ to  atomic transition frequencies (these contributions produce mass and field isotope shifts). These contributions are  small when compared to the optical transition frequency; however, this mechanism indicates a sensitivity to the parameter $X_{e} = m_{e}/\Lambda_{\text{QCD}}$ from measurements comparing the ratio of two optical transition frequencies.
In general, comparing two transition frequencies $\nu_{a}$ and $\nu_{b}$, we obtain

\begin{align} \label{variation}
    \frac{\delta (\nu_{a} / \nu_{b})}{(\nu_{a} / \nu_{b})} = (K_{1} - K_{2}) \frac{\delta r_{N}^{2}}{r_{N}^{2}}  + (K_{3} - K_{4}) \frac{\delta m_{N}}{m_{N}} \,,
\end{align}
where

\begin{align}
    K_{1} & = \frac{K_{\text{FS}}^{\nu_{a}} r_{N,a}^{2} }{\nu_{a}} \,, \\
    K_{2} & = \frac{K_{\text{FS}}^{\nu_{b}} r_{N,b}^{2} }{\nu_{b}} \,, \\
    K_{3} & =   \frac{K_{\text{MS}}^{\nu_{a}}}{\nu_{a} m_{N,a}} \,, \\
    K_{4} & = \frac{K_{\text{MS}}^{\nu_{b}}}{\nu_{b} m_{N,b}} \,,
\end{align}
where $K_{\text{MS}}$ and $K_{\text{FS}}$ are the conventional mass and field shift coefficients respectively and $m_{N}$ is the mass of the nucleus. 
The mass shift term dominates for light nuclei whilst the field shift term dominates for heavy nuclei.

Before proceeding to specific numbers, we discuss what dimensionless parameters are involved in these variations. These effects are extracted from the measurement of the variation of the ratio of two atomic transition frequencies which have different dependencies on varying parameters. In this case, one may say that the dimensionless parameters appear after dividing the variations of dimensionful parameters by the appropriate atomic units of energy $E_{h} = \alpha^{2} m_{e} c^{2}$, length $a_B=\hbar /(m c \alpha)$,  etc.

The nucleon mass dependence in light nuclei originates from the dependence of optical transition frequencies on the reduced mass \footnote{This equation accounts for the normal mass shift only, the specific mass shift depends on the transition  and is omitted here.} 
\begin{align}
    \bar{\mu} = \frac{m_{e} m_{N}}{m_{e} + m_{N}} \,,
\end{align}
where the nuclear mass $m_{N}$ is proportional to the nucleon mass, $m_{N} \approx A m_{n}$. Therefore, the dimensionless parameter here is  $X_{e} = m_e/m_n \approx  m_{e}/\Lambda_{\text{QCD}}$.

The sensitivity to other hadronic parameters comes about from the energy shift due to a finite nuclear size. In high-mass nuclei, the relative change in energy (or electronic transition frequency $\nu$) in atoms is proportional to 

\begin{align}
   \frac{\delta E}{E} \propto \left( \frac{r_{N}}{a_{B}} \right)^{2} = \left( \frac{r_{N} m_{e} \alpha c}{\hbar} \right)^{2} \,.
\end{align}
Setting $r_{N} = A^{1/3} r_{0}$, we yield
\begin{align} \label{rNaB}
    \left( \frac{r_{N}}{a_{B}} \right)^{2} =  \left( r_{0} A^{1/3} m_{e} \alpha \right)^{2} \,.
\end{align}
The variation of $r_{N}^{2}$ is proportional to {(see Eqns (\ref{rNaB}) and (\ref{deltar0Lambda}))

\begin{align}\label{rLambda}
    \frac{\delta (r_{N}/a_B)^{2}}{(r_{N}/a_B)^{2}} & = 2 \left( \frac{\delta m_{e}}{m_{e}} + \frac{\delta \alpha}{\alpha}  - 3.5 \frac{\delta \Lambda_{\text{QCD}}}{\Lambda_{\text{QCD}}} + 0.3 \frac{\delta m_{q}}{m_{q}} \right) \,.
\end{align}

Now we discuss specific numerical results which have been derived from the ratio of the optical transition frequencies of $^{1}S_{0} \rightarrow \ ^{3}P_{0}$ transitions in Al$^{+}$ and $^{2}S_{1/2} \rightarrow \ ^{2}D_{5/2}$ transitions in Hg$^{+}$~\cite{Wineland2007,AlHgdrift,Tang2021,DzubaFlambaumMansour2024} and the ratio of the $^{2}S_{1/2} (F = 0) \leftrightarrow \ ^{2} F_{7/2} (F=3)$ electric octupole (E3) and $^{2}S_{1/2} (F = 0) \leftrightarrow \ ^{2} D_{3/2} (F=2)$ electric quadrupole (E2) transition frequencies in Yb$^{+}$~\cite{Filzinger2023,Banerjee2023}. In our papers Refs.~\cite{FlambaumMansour2023,DzubaFlambaumMansour2024} we used these results to obtain constrains on variation of nucleon mass, nuclear radius, and other hadron parameters. 
The constraints 
presented in these references and Eq. (\ref{rLambda}) may be repurposed into constraints on $\Lambda_{\text{QCD}}$. These constraints are presented in Table \ref{tab:qcd_variation}.

\begin{table}[h]
\centering
\resizebox{\columnwidth}{!}{%
\begin{tabular}{ccccc}
\hline
Clock 1 & Clock 2 & Constraint (yr$^{-1}$) & Ref. & $\frac{\delta \Lambda_{\text{QCD}}}{\Lambda_{\text{QCD}}}$ (yr$^{-1}$)\\
\hline
\hline
Al$^+$ & Hg$^+$ & $(-5.3 \pm 7.9) \times 10^{-17}$ & \cite{AlHgdrift} & $ (3.2 \pm 4.9) \times 10^{-15}$ \\

Yb$^+$(E3) & Yb$^+$(E2) & $(-1.2 \pm 1.8) \times 10^{-18}$ & \cite{Filzinger2023} &  $ (7.2 \pm 11) \times 10^{-17}$ \\
\hline
\end{tabular}
}
\caption{Constraints on fractional variation of $\Lambda_{\text{QCD}}$ from the comparison of two optical clock transition frequencies in different systems.}
\label{tab:qcd_variation}
\end{table}

\section{Sensitivity of  the  $^{229}$Th based nuclear clock to $\Lambda_{\text{QCD}}$ }

Refs.~\cite{Wiringa2007,Wiringa2009} have calculated the sensitivity of the binding energy and nuclear radius in various light nuclei to the different hadron masses. These calculations were extrapolated to heavier nuclei.  In such nuclei, the errors produced by such extrapolation were found to be smaller than the errors of direct calculations in heavy nuclei.  Assuming that $\Lambda_{\text{QCD}}$ does not vary, these calculations were  used to determine that the effects of the variation of the 
quark mass $m_q$ are enhanced roughly $ 10^{5}$ times in the $ 8.4 $ eV ``nuclear clock'' transition between the ground and first excited states in $^{229}$Th and about $ 10^{8}$ times in the relative shift of the 0.1 eV compound resonance in $^{150}$Sm.

Let us now investigate the implications of interpreting their results as being due to a variation of $\Lambda_{\text{QCD}}$. Using Table III of Ref.~\cite{Wiringa2009}, the variation of the binding energy $E_{S}$ is related to the variation of the hadron masses by \footnote{The contribution of the electromagnetic interaction will be considered  below.}




\begin{align} \label{BindingEnergy}
    \frac{\delta E_{S}}{E_{S}} \approx \Delta \mathcal{E}(m_{N}) \frac{\delta m_{N}}{m_{N}}  + \Delta \mathcal{E}(m_{\pi}) \frac{\delta m_{\pi}}{m_{\pi}} \,,
\end{align}
where $m_{N,\pi}$ represent the nucleon and pion masses and $\Delta \mathcal{E}(m_{N, \pi})$ represent their respective dimensionless derivatives

\begin{align}
    \Delta \mathcal{E}(m_{N,\pi}) =   \frac{\delta E_{S}/E_{S} }{\delta m_{N,\pi} / m_{N,\pi} } \,,
\end{align}
which were roughly calculated to be $\Delta \mathcal{E}(m_{N}) \approx 12$ and $\Delta \mathcal{E}(m_{\pi}) \approx -6$. Using Eqs. (\ref{LambdaQCDFC}) and (\ref{GMOR}), we yield the following result for the variation of the binding energy $E_{S}$





\begin{align} \label{Ebvariation}
    \frac{\delta E_{S}}{E_{S}} \approx 7.9 \frac{\delta \Lambda_{\text{QCD}}}{\Lambda_{\text{QCD}}} - 1.5 \frac{\delta m_{q}}{m_{q}} \,.
\end{align}


This expression indicates that the variation of the binding energy $E_{S}$ seems to be $\sim 5$ times more sensitive to variations of $\Lambda_{\text{QCD}}$ than to variations of the average light quark mass $m_{q}$.

 Note that the binding energy $E_S$ includes contributions from the strong interaction and the kinetic energy. The result for $\delta E_{S}/E_{S}$  was extrapolated from calculations performed for light nuclei, since the binding energy per nucleon is approximately the same in light and heavy nuclei. However, this expression does not include the contribution of the electromagnetic energy $E_C$ which is small in light nuclei ($E_C \propto Z (Z-1)$).


These results have implications for experiments with  the ``nuclear clock'' transition between the ground and first excited states of the $^{229}$Th nucleus, first  proposed in Ref.~\cite{Peik2003}, where effects of the variation of fundamental constants are strongly enhanced \cite{FlambaumTh2006}.
This transition has been measured by a range of experiments, see e.g.~\cite{Seiferle2019,Masuda2019,Yamaguchi2019,Sikorski2020,Katori2024,ZhangBao2024,Tiedau2024,Elwell2024,Hiraki2024,Zhang2024}. The most precise measurement gives a transition frequency of $\omega = 2,020,407,384,335 \ (2)$ kHz ( 8.36 eV) for $^{229}$Th dopant ions embedded in a calcium fluoride crystal~\cite{Zhang2024}. 

Using the results presented above, we may write this transition frequency $\omega$ as a sum of nearly canceling components that have different dependencies on the fundamental constants. The low-lying isomeric state in $^{229}$Th is theorised to exhibit a different shape and spin configuration to the ground state~\cite{Litvinova,Minkov2021}, resulting in a small energy shift due to the residual strong and electromagnetic effects. As such, we may write $\omega$ as a sum of the electromagnetic energy contribution ($E_C$) and strong + kinetic energies contribution $E_S$:

\begin{align} \label{sumofenergies}
    \omega = E_{C} + E_{S} = 8.36 \ \text{eV} \,.
\end{align}


The sensitivity of the nuclear transition energy to changes in the fine structure constant $\alpha$ is 

\begin{align}
    \frac{\delta \omega}{\omega} = K_{\alpha} \frac{\delta \alpha}{\alpha} \,,
\end{align}
where $K_{\alpha}$ may be determined by the relation~\cite{FlambaumTh2006}

\begin{align}
    K_{\alpha} = \frac{E_{C}}{\omega} \,.
\end{align}
In Ref.~\cite{PhysRevLett.102.210801} the sensitivity coefficient $K_{\alpha}$ was expressed via the difference in squared nuclear radii $\delta r^2$ and the difference of the electric quadrupole moments $\delta Q$ between the ground and isomeric states. \footnote{Due to the lack of accurate data for electric quadrupole moments,  Ref.~\cite{Fadeev2020} used the approximation of a constant nuclear density to express $\delta Q$ via a measured value of $\delta r^2$ and obtained $K_{\alpha}$ = $-0.82 \ (0.25) \times 10^{4}$, with only the experimental error being presented. However, the approximation of a constant nuclear density is not sufficiently accurate in the case where the contributions of $\delta r^2$ and $\delta Q$ have opposing signs.} 
Recent, 
measurements of $\delta r^2$ and $\delta Q$~\cite{SafronovaYe}, using the formula expressing $K_{\alpha}$ in terms of these parameters~\cite{PhysRevLett.102.210801}, give $K_{\alpha}$ = $0.59 \ (0.23) \times 10^{4}$ \footnote{Ref. ~\cite{SafronovaYe} also investigated the sensitivity of $K_{\alpha}$ to a possible octupole deformation.}
.  

We may use this information, along with the results presented above, in order to estimate the nuclear transition's sensitivity to changes in the QCD scale $\Lambda_{\text{QCD}}$. In similar fashion to the analysis performed in Refs.~\cite{FlambaumTh2006,Wiringa2009}, Equation (\ref{sumofenergies}) implies $ E_{C} \approx -E_{S}$ and 

\begin{align}\label{omega}
    \frac{\delta \omega}{\omega} \approx \frac{E_{C}}{\omega} \left( \frac{\delta E_{C}}{E_{C}} - \frac{\delta E_{S}}{E_{S}} \right)\,.
\end{align}
The Coulomb interaction contribution $E_C$ is proportional to $e^2/r_0 \propto {\alpha} \Lambda_{\text{QCD}}$, see Eq. (\ref{rLambda}).

Using Eq. (\ref{omega}), along with the sensitivity of $E_{S}$ to $\Lambda_{\text{QCD}}$ presented in (\ref{Ebvariation}), we yield

\begin{align} 
\begin{split} \label{ThFrequency}
   \frac{\delta \omega}{\omega} & \approx K_{\alpha} \left( \frac{\delta \alpha}{\alpha} - 7 \frac{\delta \Lambda_{\text{QCD}}}{\Lambda_{\text{QCD}}} + 1.5 \frac{\delta m_{q}}{m_{q}} \right) \\
   & \approx 0.6 \times 10^{4}\left( \frac{\delta \alpha}{\alpha} - 7 \frac{\delta \Lambda_{\text{QCD}}}{\Lambda_{\text{QCD}}} + 1.5 \frac{\delta m_{q}}{m_{q}} \right)\,,
\end{split}
\end{align}
which  implies that the nuclear clock transition is also more sensitive to variations in $\Lambda_{\text{QCD}}$. 

This estimate for the sensitivity of the nuclear clock transition to variations of $\Lambda_{\text{QCD}}$ may have implications for the measurements  performed in Ref.~\cite{Fuchs2024PRX}, which present constraints on the fractional variation of $\Lambda_{\text{QCD}}$ due to an interaction with ultralight dark matter, taking a  rough estimate of the sensitivity coefficient as $K_{\text{QCD}} \sim K_{\alpha} \sim 10^4 - 10^5$. Eq. (\ref{ThFrequency}) allows one to refine such constraints, which are expected to be significantly improved in future measurements.




This effect is supposed to be extracted from the measurement of the variation of the dimensionless ratio of the nuclear transition frequency to an atomic transition frequency, which is proportional to the atomic unit of energy $E_h$.  For example, in Ref.~\cite{Zhang2024} the ratio of the nuclear transition frequency to an electron transition frequency in the Sr atom has been measured. The relative variation of the Sr transition frequency is not enhanced, so we may neglect its variation. 

\section{Constraint from Big Bang Nucleosynthesis}

Big-Bang Nucleosynthesis (BBN) took place during the first few minutes after the Big Bang, when the light nuclides H, D, He and Li were produced. Their primordial abundances are exquisitely sensitive to nuclear binding energies - and hence to the hadronic parameters that set those bindings. By comparing the observed primordial abundances with BBN calculations one can therefore bound any change in the underlying constants since the early Universe.

Ref.~\cite{Wiringa2007} (see previous estimates in Refs. \cite{Shuryak2002,FlambaumShuryak,Dmitriev2003,Dmitriev2004} and also~\cite{BERENGUT2010114}) computed how the binding energies in light nuclei respond to shifts of the hadron masses and, assuming the QCD scale $\Lambda_{\text{QCD}}$ is fixed, translated the difference between calculated and observed abundances into a fractional change of the average light-quark mass:  $\delta m_{q}/m_{q} = K (0.013 \pm 0.002)$,  where $K \sim 1$ reflects a theoretical uncertainty of about a factor of 2.





Here we reinterpret that result in the complementary scenario in which the variation is driven instead by a gluon–scalar coupling that makes $\Lambda_{\text{QCD}}$ drift, while $m_q$ is held fixed.
Using the nuclear binding energy response in Eq. (\ref{Ebvariation}), we obtain 

\begin{align}
    \frac{\delta \Lambda_{\text{QCD}}}{\Lambda_{\text{QCD}}} = K (-2.5 \pm 0.4) \times 10^{-3} \,.
\end{align}

The fact that the central value is non-zero reflects the well-known ``lithium problem'' - the predicted $^{7}$Li abundance exceeds the observed abundance by a factor of 3 to 4.  It seems that the lithium anomaly survives every Standard model update to date, and the variation of $\Lambda_{\text{QCD}}$ may solve this problem. The most credible fix within the Standard model is gradual Li destruction in old stars, supported by modern diffusion-mixing simulations but not yet universally accepted. Reviews of the primordial lithium problem - its observational status, nuclear physics inputs and proposed solutions - can be found in Refs.~\cite{Fields2011,Cyburt2016}.

    

\section{Constraint from the Oklo natural nuclear reactor}
The Oklo natural nuclear reactor was a naturally occurring fission reactor that operated about 1.8 billion years ago. Its sustained chain reactions produced a neutron spectrum that covered a very low energy resonance in $^{149}$Sm. This resonance, with $E_r = 0.1\ \mathrm{eV}$, is small compared to the depth of the neutron potential ($U = -50\ \mathrm{MeV}$), giving the system a high sensitivity to variations of the fundamental constants.
Refs.~\cite{Shlyakhter1976,DAMOUR1996,FUJII2000,Gould2006,Petrov2006} used this high sensitivity to find limits on the variation of $\alpha$, while Refs.~\cite{Shuryak2002,FlambaumShuryak,Wiringa2009} estimated limits on variation of the quark mass.

We can express the neutron capture resonance $E_r$ as 
\begin{align}
\begin{split}
E_r & = E_k +U, \\
E_k & = C_1 \Lambda_{\text{QCD}} +C_2 {\alpha} \Lambda_{\text{QCD}} + C_3 m_q, \\
U & = C_4 \Lambda_{\text{QCD}} +C_5 {\alpha} \Lambda_{\text{QCD}} + C_6 m_q, 
\end{split}
\end{align}
where $E_k$ is the kinetic energy of the captured neutron inside the nucleus and $C_i $ are dimensionless parameters. The contribution of the Coulomb interaction is proportional to $e^2/r_0 \propto {\alpha} \Lambda_{\text{QCD}}$.
Substituting the expressions for $E_{k}$ and $U$ into the resonance energy gives 
\begin{align} \label{Er} 
E_r= (C_1 +C_4) \Lambda_{\text{QCD}} + (C_2+C_5)  {\alpha} \Lambda_{\text{QCD}} + (C_3 +C_6) m_q \,.
\end{align}
Now, the condition $E_{k} = -U + 0.1$eV implies that $E_{k}$ and $U$ are almost equal in magnitude. However, this near cancellation does not mean that the sensitivity of $E_r$ to the variation of fundamental constants is small, as the fact that $E_r\approx 0$ allows for all coefficients in Eq. (\ref{Er}) to be large, due to the fact that $E_k$ and $U$ are large.




Note that if only one parameter was involved (e.g. if we neglect the small contributions to the energy from $\alpha$ and $m_q$), the sensitivity to the variation of $\Lambda_{\text{QCD}}$ disappears, since $E_k$ and $U$ in this case are proportional ($E_k= C U$) and subsequently there will be no enhancement of the variation effect in their sum: $\delta E_r /E_r = \delta (E_k + U)/(E_k + U) = \delta E_k/E_k $. Another way one may consider this point is to note that the condition $E_{r} \approx 0$ would require $C_1 +C_4 \approx 0$ in the case that $m_q=\alpha=0$. As a result, the dependence on $\Lambda_{\text{QCD}}$ is eliminated in accordance with Eq. (\ref{Er}).


The experimental condition $E_r \approx 0$ allows us to express $C_1 +C_4$ in terms of other constants. Doing so, we obtain 
 \begin{align} \label{Er0}
 \begin{split}
E_r & =  - \frac{\Lambda_{\text{QCD}}}{\Lambda^0_{\text{QCD}}} \bigg[ (C_2+C_5){\alpha^0} \Lambda^0_{\text{QCD}}+(C_3 +C_6) m^0_q \bigg] \\
& + (C_2+C_5)  {\alpha} \Lambda_{\text{QCD}} + (C_3 +C_6) m_q \,.
\end{split}
\end{align}
where $\Lambda^0_{\text{QCD}}$, $\alpha^0$ and $m^0_q$ are the present values of these constants. In doing so, we obtain the following for the variation of the resonance energy

\begin{align} \label{ErS}
\begin{split}
\delta E_r & = - \frac{\delta \Lambda_{\text{QCD}}}{\Lambda^0_{\text{QCD}}}\bigg[ (C_2+C_5){\alpha^0} \Lambda^0_{\text{QCD}}+(C_3 +C_6) m^0_q \bigg] \\
& + (C_2+C_5) \delta ( \alpha\Lambda_{\text{QCD}}) + (C_3 +C_6) \delta m_q \,.
\end{split}
\end{align}
The shift of the resonance due to the variation of $\alpha$ was calculated in Refs.~\cite{Shlyakhter1976,DAMOUR1996} to be 
\begin{align}
    \delta E_{r} = (1.1 \pm 0.1) \frac{\delta \alpha}{\alpha} \ \text{MeV} \,,
\end{align}
while the shift  due to the variation of $m_{q}$  

\begin{align}
    \delta E_{r} \approx  10 \frac{\delta m_{q}}{m_{q}} \ \text{MeV} \,,
\end{align}
was calculated in Ref.~\cite{Wiringa2009}. Using these results, we obtain

\begin{align} \label{ErMev}
\begin{split}
\delta E_r & \approx  10 \, {\text{MeV}}\, \left(- 1.1 \frac{\delta \Lambda_{\text{QCD}}}{\Lambda_{\text{QCD}}}+ \frac{\delta m_q} {m_q} - 0.1 \frac{\delta \alpha} {\alpha} \right) \,, \\
& \approx  10 \, {\text{MeV}}\, \left( \frac{\delta X_q} {X_q} - 0.1 \frac{\delta \alpha} {\alpha} \right) \,.
\end{split}
\end{align}

Now we should specify how the shift of $E_r$ has been measured~\cite{Shlyakhter1976,DAMOUR1996}. It was expressed through the measured ratio of abundances of  $^{149}$Sm and $^{150}$Sm near the Oklo natural nuclear reactor and compared to this ratio in other areas where there was no neutron flux transforming $^{149}$Sm to $^{150}$Sm. The neutron flux in Oklo was produced by the nuclear fission of $^{235}$U, and the integrated neutron flux was found by comparing the difference between the abundance of $^{235}$U in Oklo and its natural abundance in other places.


The observed shift in the resonance energy $\delta E_r$ between the present and time of the Oklo natural reactor 1.8 billion years ago was constrained to be less than 0.02 eV~\cite{Gould2006,Petrov2006,FUJII2000}, which provides the constraint
\begin{align} \label{ErLimitXq} 
\bigg| \frac{\delta X_q} {X_q} - 0.1 \frac{\delta \alpha} {\alpha} \bigg| < 2 \times 10^{-9}\,.
\end{align}
The variations of $\alpha$, $m_{q}$ and $\Lambda_{\text{QCD}}$ are produced by different terms in the interaction of dark matter or dark energy fields with gluons, quarks and photons. If we assume independent variations of these constants, we obtain limits on variation of these parameters during the 1.8 billion years 
\begin{align} \label{ErLimit} 
\bigg| \frac{\delta \alpha} {\alpha} \bigg| & < 2 \times 10^{-8} \,, \\
\bigg| \frac{\delta m_q} {m_q} \bigg| & < 2 \times 10^{-9}\,,\\
\bigg| \frac{\delta \Lambda_{\text{QCD}}}{\Lambda_{\text{QCD}}} \bigg| & < 2 \times 10^{-9} \,, 
\end{align}
which may be compared to atomic clock results, by assuming the linear drift of these parameters:   
\begin{align} \label{Erdrift} 
\bigg| \frac{\delta \alpha} {\alpha} \bigg| < 1 \times 10^{-17} \ \text{yr}^{-1} \,, \\
\bigg| \frac{\delta m_q} {m_q} \bigg| < 1 \times 10^{-18} \ \text{yr}^{-1}\,, \\  
\bigg| \frac{\delta \Lambda_{\text{QCD}}}{\Lambda_{\text{QCD}}} \bigg|< 1 \times 10^{-18} \ \text{yr}^{-1}\,.
\end{align}
This chain of arguments is possible since the fission properties of $^{235}$U have a relatively low sensitivity to the variation of fundamental constants (compared to $^{149}$Sm).


This result can be interpreted in the context of dark-energy-type models \cite{BarrowMagueijo1998,Barrow2005,BarrowSandvikMagueijo2002,Wetterich2003,Uzan}, where variations occur on cosmological scales. Another class of models involves dark halos, in which the dark matter density around massive bodies is strongly enhanced compared to the average dark matter density and evolves on timescales of billions of years (see, e.g., \cite{BanerjeeFlambaum2020, Banerjee2020, Budker2023}). Dynamical mechanisms of this type predict present-day fractional variations at the level of $10^{-13}$–$10^{-16}$~\cite{Fuchs2024PRX}, far below the $10^{-9}$ sensitivity accessible with Oklo. Therefore, Oklo may only probe an exceptionally large increase of the dark matter density. Large variations in the dark matter density could also accur due to passing dark matter clouds or Bose stars. In this context, Oklo constraints on $\Lambda_{\text{QCD}}$ may rule out the possibility that such extreme local enhancements of the dark matter density occurred in the past or are occurring today.

Now, we discuss the possible inconsistencies of this approach and seek a clarification.  The results are based on the measurements of the dimensionless abundances, meaning they may be expressed in terms of dimensionless ratios of the fundamental constants, without the use of any ``human-introduced'' units like MeV (which also may be subject to variation) for the Coulomb and strong interaction's contributions to energy, see Eq. (\ref{ErMev}).  The key point is that we are interested in a resonance effect enhanced by a factor of $10^{8}$. As such, the choice of units is not important, as variations in the ratios of different units are not enhanced.


We still may pose a purely theoretical question, is it possible to avoid using ``human-introduced'' units and measure the effect in natural units for this problem from the start?  The origin of the MeV coefficient in the expressions for the Coulomb and strong interaction  energies can  be traced back to $\Lambda_{\text{QCD}}$, which mainly determines nuclear properties. For example, the Coulomb interaction in nuclei is proportional to $e^2/r_0 \sim \alpha \Lambda_{\text{QCD}}$. Therefore, we may make the substitution 10 MeV $\to \Lambda_{\text{QCD}}/30$. 
The change in the abundance of $^{149}$Sm is proportional to the change in the abundance of $^{235}$U, which produced the neutron flux captured by $^{149}$Sm. The proportionality coefficient includes the ratio of the resonance neutron capture cross section to a geometric factor, which is proportional to the squared internuclear distance in solids, approximately $a_B^2= 1/(\alpha m_e)^2$. Following this line of reasoning, we may conclude that the relevant dimensionless parameter is $E_r/(\alpha m_e)$.

Further analysis, including addressing the neutron cooling process needed for an efficient neutron capture, involves a dependence on other fundamental physical constants such as Newton's gravitational constant $G_N$, to which the Earth's temperature is sensitive. However, these gravitational and other non-resonant effects are not enhanced and may be ignored.

\section{Constraints from Quasar Absorption Spectra}
Quasar absorption spectra provide an extremely sensitive astrophysical probe of the variation of fundamental constants over cosmological timescales - see e.g \cite{BarrowMagueijo1998,Barrow2005,BarrowSandvikMagueijo2002,Wetterich2003,Uzan,Murphy2001,MurphyWebbFlambaum2003,Reinhold2006,King2008,Thompson2009,Malec2009,Murphy2008,Menten2008,Henkel2009,Petitjean2009,Kanekar2005,Darling2004,Tzanavaris2007,Srianand2010,Levshakov2008,Chengalur2003,Kanekar2010}. Intervening clouds along the line of sight of the quasar give access to the spectra of the atoms and ions present in the cloud. Calculations of the dependence of corresponding atomic transition frequencies on the fundamental constants have been performed e.g. in Refs. \cite{DZUBAPRL1999,DZUBAPRA1999,Thomas,Tedesco,CanJPhQ,Ammonia,KozlovMolecules} and references therein.  Through the analysis of atomic and molecular transitions in such absorption systems, measurements of the variation of the fine structure constant $\alpha$, the proton to electron mass ratio $\mu$ and composite parameters $F \equiv g_{p} (\alpha^{2} \mu)^{1.57}$, $x \equiv \alpha^{2} g_{p}/\mu$, $F^{\prime} \equiv \alpha^{2} \mu$ and $G \equiv g_{p} (\alpha \mu)^{1.85}$ have been performed, where $g_{p}$ is the proton $g$-factor; see e.g. Refs.~\cite{Reinhold2006,King2008,Thompson2009,Malec2009,Murphy2008,Menten2008,Henkel2009,Petitjean2009,Kanekar2005,Darling2004,Tzanavaris2007,Srianand2010,Levshakov2008,Chengalur2003,Kanekar2010}.

Aside from $\alpha$, these parameters are all sensitive to changes in $\Lambda_{\text{QCD}}$. In this section, we re-interpret constraints on the variation of these $\Lambda_{\text{QCD}}$-dependent parameters from such measurements as limits on the variation of $\Lambda_{\text{QCD}}$. We once again do so under the assumption that this parameter acquires a time dependence upon interaction with an ultralight dark matter or dark energy field which couples to the gluon field, leaving $\alpha$ and the electron mass unchanged.  

The proton mass $m_p$ and magnetic $g$-factor $g_p$ both depend  on $\Lambda_{\text{QCD}}$. Within  10\% accuracy, $\frac{\delta \mu}{\mu} \approx \frac{\delta \Lambda_{\text{QCD}}}{\Lambda_{\text{QCD}}}$ - see Eq. (\ref{LambdaQCDFC}). A detailed calculation of the quark mass dependence of $g_{p}$ in Ref.~\cite{Thomas} together with Eqs. (\ref{muN},\ref{muN1}) leads to the estimate $\frac{\delta g_p}{g_p} \approx 0.10 \frac{\delta \Lambda_{\text{QCD}}}{\Lambda_{\text{QCD}}}$. 
This allows us to translate existing astrophysical bounds, as compiled in Ref.~\cite{Uzan}, into direct constraints on $\Lambda_{\text{QCD}}$, see Table \ref{QASConstraintsTable}. Under the assumption of a linear variation of $\Lambda_{\text{QCD}}$, these constraints are remarkably close to those obtained from atomic clock measurements. We note, however, that this assumption has no justification over such an extensive period of time.

\section{Summary}

Some theories beyond the Standard model predict that “fundamental constants’’ might drift (or oscillate) with time or space. Only dimensionless constants - such as the fine-structure constant $\alpha$ can be constrained in a  model-independent way: any change in a pure number is unambiguous. By contrast, limits on dimensionful quantities ($c$, $\hbar$, $G_N$, $\Lambda_{\text{QCD}}$, $m_e$,…) make sense only within a specified framework - for example, a model where an ultralight dark-matter or dark-energy field couples to ordinary matter. The reason is simple: the “human’’ units used to quote those quantities (seconds, metres, electron-volts, Hertz, …) are themselves defined through physical processes that depend on the underlying constants. Thus, the units may vary too and their dependence on the fundamental constants may be quite complicated. For example, the dependence of the hyperfine transition frequency in Cs, which defines the second and the Hertz, can be seen in Table \ref{Clockvariations}. Meaningful constraints on dimensionful parameters therefore require stating how the units are anchored in the chosen model; otherwise one can always absorb an apparent variation by redefining the ruler or the clock.

\begin{widetext}

\begin{table}[h!]
\centering
\begin{tabular}{ccccccc}
\toprule
\multirow{2}{*}{Constant} & Constraint & \multirow{2}{*}{$z$} & \multirow{2}{*}{Ref.} & Lookback Time & $\delta\Lambda_{\text{QCD}}/\Lambda_{\text{QCD}}$ & $\delta\Lambda_{\text{QCD}}/\Lambda_{\text{QCD}}$ \\
& \((\times 10^{-5})\) & & & \((\times 10^{10} \ \text{yr})\) & ($\times 10^{-5}$) & \((\times 10^{-15}\ \text{yr}^{-1})\) \\
\midrule
\multirow{11}{*}{$\mu$} 
& $(2.78 \pm 0.88)$ & 2.59 & \cite{Reinhold2006} & 1.12 & \((2.78 \pm 0.88)\) & \(2.49 \pm 0.79\) \\
& $(2.06 \pm 0.79)$ & 3.02 & \cite{Reinhold2006} & 1.13 & \((2.06 \pm 0.79)\) & \(1.78 \pm 0.68\) \\
& $(1.01 \pm 0.62)$ & 2.59 & \cite{King2008} & 1.12 & \((1.01 \pm 0.62)\) & \(0.90 \pm 0.55\) \\
& $(0.82 \pm 0.74)$ & 2.8  & \cite{King2008} & 1.14 & \((0.82 \pm 0.74)\) & \(0.72 \pm 0.65\) \\
& $(0.26 \pm 0.30)$ & 3.02 & \cite{King2008} & 1.16 & \((0.26 \pm 0.30)\) & \(0.22 \pm 0.26\) \\
& $(0.70 \pm 0.80)$ & 3.02 & \cite{Thompson2009} & 1.16 & \((0.70 \pm 0.80)\) & \(0.61 \pm 0.69\) \\
& $(0.56 \pm 0.55_{\text{stat}} \pm 0.27_{\text{syst}})$ & 2.06 & \cite{Malec2009} & 1.05 & \((0.56 \pm 0.61)\) & \(0.53 \pm 0.58\) \\
& $< 0.18$ & 0.685 & \cite{Murphy2008} & 0.629 & $< 0.18$ & $< 0.29$ \\
& $< 0.38$ & 0.685 & \cite{Menten2008} & 0.629 & $< 0.38$ & $< 0.60$ \\
& $< 0.14$ & 0.89 & \cite{Henkel2009} & 0.638 & $< 0.14$ & $< 0.22$ \\
& $< 9$ & 2.42 & \cite{Petitjean2009} & 1.10 & $< 9$ & $< 8.2$ \\
\hline

\multirow{2}{*}{$F$} 
& $(-0.44 \pm 0.36 \pm 1.0_{\text{syst}})$ & 0.765 & \cite{Kanekar2005} & 0.673 & \((-0.28 \pm 0.68)\) & \(-0.42 \pm 1.0\) \\
& $(0.51 \pm 1.26)$ & 0.247 & \cite{Darling2004} & 0.293 & \((0.33 \pm 0.80)\) & \(1.11 \pm 2.74\) \\
\hline

\multirow{2}{*}{$x$} 
& $(-0.63 \pm 0.99)$ & 1.29 & \cite{Tzanavaris2007} & 0.883 & \((0.63 \pm 0.99)\) & \(0.71 \pm 1.12\) \\
& $(-0.17 \pm 0.17)$ & 3.17 & \cite{Srianand2010} & 1.17 & \((0.17 \pm 0.17)\) & \(0.15 \pm 0.15\) \\
\hline

\multirow{2}{*}{$F^{\prime}$} 
& $(1.0 \pm 10)$ & 4.69 & \cite{Levshakov2008} & 1.24 & \((1.00 \pm 10.00)\) & \(0.80 \pm 8.0\) \\
& $(14 \pm 15)$ & 6.42 & \cite{Levshakov2008} & 1.29 & \((14 \pm 15)\) & \(11 \pm 12\) \\
\hline

\multirow{3}{*}{$G$} 
& $< 1.1$ & 0.247 & \cite{Chengalur2003} & 0.294 & \(< 0.60\) & \(< 2.03\) \\
& $< 1.16$ & 0.0018 & \cite{Chengalur2003} & 0.0025 & \(< 0.63\) & \(< 250\) \\
& $(-1.18 \pm 0.46)$ & 0.247 & \cite{Kanekar2010} & 0.294 & \((-0.64 \pm 0.25)\) & \(-2.17 \pm 0.85\) \\
\bottomrule
\end{tabular}
\caption{Constraints on the variation of $\Lambda_{\text{QCD}}$ from Quasar Absorption Spectra. The constants (as presented in Ref.~\cite{Uzan}) are: $\mu \equiv m_{p}/m_{e}$, $F \equiv g_{p} (\alpha^{2} \mu)^{1.57}$, $x \equiv \alpha^{2} g_{p}/\mu$, $F^{\prime} \equiv \alpha^{2} \mu$ and $G \equiv g_{p} (\alpha \mu)^{1.85}$.}
\label{QASConstraintsTable}
\end{table}

\end{widetext}

To consider the variation of dimensionful constants one should specify an interaction which produces this variation.
In this paper, we have investigated the sensitivity of several  systems to the variation of the dimensionful QCD interaction constant $\Lambda_{\text{QCD}}$ and extracted limits on its drift. In particular, we consider the case in which an interaction between an evolving dark matter or dark energy scalar field $\phi$ and the gluon field induces a time dependence in $\Lambda_{\text{QCD}}$, while leaving $\alpha$ and the fundamental  masses unchanged.


Within the minimal scenario in which this gluonic channel is the sole $\phi$ interaction, we show that the atomic Yb/Cs  clock comparison \cite{YbCs} places the constraint $\dot{\Lambda}_{\text{QCD}}/\Lambda_{\text{QCD}}=(3.2 \pm 3.5) \times 10^{-17} \ \text{yr}^{-1}$. Further independent constraints are obtained from isotopic ratios measured in the Oklo natural nuclear reactor, which limit the variation to $|\delta\Lambda_{\text{QCD}}/\Lambda_{\text{QCD}}|<2\times10^{-9}$ over 1.8 billion years. Assuming a linear drift, this translates to a bound  $|\dot{\Lambda}_{\text{QCD}}/\Lambda_{\text{QCD}}|<1\times10^{-18} \text{yr}^{-1}$. Additionally, the sensitivity of the $^{229}$Th nuclear clock to the variation of $\Lambda_{\text{QCD}}$ is estimated to be four orders of magnitude higher than that of current atomic clocks. 

We have also obtained constraints on $\Lambda_{\text{QCD}}$ variation from  quasar absorption spectra (see Table \ref{QASConstraintsTable}) and Big Bang Nucleosynthesis data. Comparison of these astrophysical and cosmological constraints with atomic clock limits requires the assumption of a linear variation of $\Lambda_{\text{QCD}}$, which has no justification over such large time intervals (from 3 to 13 billion years). However, if a linear variation is assumed, limits derived from the quasar absorption spectra are surprisingly close to those from atomic clocks.

The BBN data indicates a non-zero deviation from the present value,  $\delta \Lambda_{\text{QCD}}/\Lambda_{\text{QCD}} = (-2.5 \pm 0.4) \times 10^{-3}$. This is a reflection of the well-known ``lithium problem,'' wherein the predicted abundance of $^{7}$Li within the Standard Model exceeds the observed abundance by a factor of 3 to 4. The variation of $\Lambda_{\text{QCD}}$ may fix this problem. However, the search for solutions within the Standard Model continues.

\section*{Acknowledgements}
We are grateful to Dima Budker, Abhishek Banerjee and Gilad Perez for valuable discussions and for bringing relevant references to our attention. This work was supported by the Australian Research Council Grant No.\ DP230101058.

\bibliographystyle{apsrev4-2}
\bibliography{References.bib}

@article{PhysRevLett.102.210801,
  title = {Proposed Experimental Method to Determine $\ensuremath{\alpha}$ Sensitivity of Splitting between Ground and 7.6 eV Isomeric States in $^{229}\mathrm{Th}$},
  author = {Berengut, J. C. and Dzuba, V. A. and Flambaum, V. V. and Porsev, S. G.},
  journal = {Phys. Rev. Lett.},
  volume = {102},
  issue = {21},
  pages = {210801},
  numpages = {4},
  year = {2009},
  month = {May},
  publisher = {American Physical Society},
  doi = {10.1103/PhysRevLett.102.210801},
  url = {https://link.aps.org/doi/10.1103/PhysRevLett.102.210801}
}

@article{FlambaumTh2006,
  title = {Enhanced Effect of Temporal Variation of the Fine Structure Constant and the Strong Interaction in $^{229}\mathrm{Th}$},
  author = {Flambaum, V. V.},
  journal = {Phys. Rev. Lett.},
  volume = {97},
  issue = {9},
  pages = {092502},
  numpages = {3},
  year = {2006},
  month = {Aug},
  publisher = {American Physical Society},
  doi = {10.1103/PhysRevLett.97.092502},
  url = {https://link.aps.org/doi/10.1103/PhysRevLett.97.092502}
}

@article{Dinh2009,
	doi = {10.1103/physreva.79.054102},
	year = 2009,
	month = {may},
	publisher = {American Physical Society ({APS})},
	volume = {79},
	number = {5},
        pages = {054102},
	author = {T. H. Dinh and A. Dunning and V. A. Dzuba and V. V. Flambaum},
	title = {Sensitivity of hyperfine structure to nuclear radius and quark mass variation},
	journal = {Phys. Rev. A}
}

@misc{Banerjee2023,
      title={Oscillating nuclear charge radii as sensors for ultralight dark matter}, 
      author={Abhishek Banerjee and Dmitry Budker and Melina Filzinger and Nils Huntemann and Gil Paz and Gilad Perez and Sergey Porsev and Marianna Safronova},
      year={2023},
      eprint={2301.10784},
      archivePrefix={arXiv},
      primaryClass={hep-ph}
}

@article{Filzinger2023,
  title = {Improved Limits on the Coupling of Ultralight Bosonic Dark Matter to Photons from Optical Atomic Clock Comparisons},
  author = {Filzinger, M. and D\"orscher, S. and Lange, R. and Klose, J. and Steinel, M. and Benkler, E. and Peik, E. and Lisdat, C. and Huntemann, N.},
  journal = {Phys. Rev. Lett.},
  volume = {130},
  issue = {25},
  pages = {253001},
  numpages = {6},
  year = {2023},
  month = {Jun},
  publisher = {American Physical Society},
  doi = {10.1103/PhysRevLett.130.253001},
  url = {https://link.aps.org/doi/10.1103/PhysRevLett.130.253001}
}

@article{Hees2016,
  title = {Searching for an Oscillating Massive Scalar Field as a Dark Matter Candidate Using Atomic Hyperfine Frequency Comparisons},
  author = {Hees, A. and Gu\'ena, J. and Abgrall, M. and Bize, S. and Wolf, P.},
  journal = {Phys. Rev. Lett.},
  volume = {117},
  issue = {6},
  pages = {061301},
  numpages = {5},
  year = {2016},
  month = {Aug},
  publisher = {American Physical Society},
  doi = {10.1103/PhysRevLett.117.061301},
  url = {https://link.aps.org/doi/10.1103/PhysRevLett.117.061301}
}

@article{Borschevsky,
    author = "Pa\v{s}teka, Luk\'a\v{s} F. and Hao, Yongliang and Borschevsky, Anastasia and Flambaum, Victor V. and Schwerdtfeger, Peter",
    title = "{Material Size Dependence on Fundamental Constants}",
    doi = "10.1103/PhysRevLett.122.160801",
    journal = "Phys. Rev. Lett.",
    volume = "122",
    number = "16",
    pages = "160801",
    year = "2019"
}

@book{Sobelman, 
address={Berlin Heidelberg New York}, 
author = {Sobelman, I.I.}, 
title={Atomic Spectra and Radiative Transitions}, 
publisher={Springer-Verlag}, 
year={1979}
}

@article{FlambaumShuryak,
  title = {Dependence of hadronic properties on quark masses and constraints on their cosmological variation},
  author = {Flambaum, V. V. and Shuryak, E. V.},
  journal = {Phys. Rev. D},
  volume = {67},
  issue = {8},
  pages = {083507},
  numpages = {7},
  year = {2003},
  month = {Apr},
  publisher = {American Physical Society},
  doi = {10.1103/PhysRevD.67.083507},
  url = {https://link.aps.org/doi/10.1103/PhysRevD.67.083507}
}

@article{Tedesco,
  title = {Dependence of nuclear magnetic moments on quark masses and limits on temporal variation of fundamental constants from atomic clock experiments},
  author = {Flambaum, V. V. and Tedesco, A. F.},
  journal = {Phys. Rev. C},
  volume = {73},
  issue = {5},
  pages = {055501},
  numpages = {9},
  year = {2006},
  month = {May},
  publisher = {American Physical Society},
  doi = {10.1103/PhysRevC.73.055501},
  url = {https://link.aps.org/doi/10.1103/PhysRevC.73.055501}
}

@article{Wiringa2007,
    author = "Flambaum, V. V. and Wiringa, Robert B.",
    title = "{Dependence of nuclear binding on hadronic mass variation}",
    doi = "10.1103/PhysRevC.76.054002",
    journal = "Phys. Rev. C",
    volume = "76",
    pages = "054002",
    year = "2007"
}

@article{Wiringa2009,
    author = "Flambaum, V. V. and Wiringa, R. B.",
    title = "{Enhanced effect of quark mass variation in Th-229 and limits from Oklo data}",
    doi = "10.1103/PhysRevC.79.034302",
    journal = "Phys. Rev. C",
    volume = "79",
    pages = "034302",
    year = "2009"
}

@article{Ammonia,
  title = {Limit on the Cosmological Variation of ${m}_{p}/{m}_{e}$ from the Inversion Spectrum of Ammonia},
  author = {Flambaum, V. V. and Kozlov, M. G.},
  journal = {Phys. Rev. Lett.},
  volume = {98},
  issue = {24},
  pages = {240801},
  numpages = {4},
  year = {2007},
  month = {Jun},
  publisher = {American Physical Society},
  doi = {10.1103/PhysRevLett.98.240801},
  url = {https://link.aps.org/doi/10.1103/PhysRevLett.98.240801}
}

@article{KozlovMolecules,
  title = {Enhanced Sensitivity to the Time Variation of the Fine-Structure Constant and ${m}_{p}/{m}_{e}$ in Diatomic Molecules},
  author = {Flambaum, V. V. and Kozlov, M. G.},
  journal = {Phys. Rev. Lett.},
  volume = {99},
  issue = {15},
  pages = {150801},
  numpages = {4},
  year = {2007},
  month = {Oct},
  publisher = {American Physical Society},
  doi = {10.1103/PhysRevLett.99.150801},
  url = {https://link.aps.org/doi/10.1103/PhysRevLett.99.150801}
}

@article{CanJPh,
    author = "Flambaum, V. V. and Dzuba, V. A.",
    title = "{Search for variation of the fundamental constants in atomic, molecular and nuclear spectra}",
    doi = "10.1139/p08-072",
    journal = "Can. J. Phys.",
    volume = "87",
    pages = "25--33",
    year = "2009"
}

@article{CanJPhQ,
    author = "Dzuba, V. A. and  Flambaum, V. V.",
    title = "{Atomic calculations and search for variation of the fine structure constant in quasar absorption spectra}",
    doi = "10.1139/p08-053", 
    journal = "Can. J. Phys.",
    volume = "87",
    pages = "15--23",
    year = "2009"
}

@article{csquarks,
    author = "Flambaum, V. V. and Munro-Laylim, P.",
    title = "{Spacetime variation of the s and c quark masses}",
    doi = "10.1103/PhysRevD.107.015004",
    journal = "Phys. Rev. D",
    volume = "107",
    number = "1",
    pages = "015004",
    year = "2023"
}

@article{PRAWebb,
  title = {Calculations of the relativistic effects in many-electron atoms and space-time variation of fundamental constants},
  author = {Dzuba, V. A. and Flambaum, V. V. and Webb, J. K.},
  journal = {Phys. Rev. A},
  volume = {59},
  issue = {1},
  pages = {230--237},
  numpages = {0},
  year = {1999},
  month = {Jan},
  publisher = {American Physical Society},
  doi = {10.1103/PhysRevA.59.230},
  url = {https://link.aps.org/doi/10.1103/PhysRevA.59.230}
}

@article{PRLWebb,
    author = "Dzuba, V. A. and Flambaum, V. V. and Webb, J. K.",
    title = "{Space-time variation of physical constants and relativistic corrections in atoms}",
    doi = "10.1103/PhysRevLett.82.888",
    journal = "Phys. Rev. Lett.",
    volume = "82",
    pages = "888--891",
    year = "1999"
}

@article{HSi,
  title = {Precision Metrology Meets Cosmology: Improved Constraints on Ultralight Dark Matter from Atom-Cavity Frequency Comparisons},
  author = {Kennedy, Colin J. and Oelker, Eric and Robinson, John M. and Bothwell, Tobias and Kedar, Dhruv and Milner, William R. and Marti, G. Edward and Derevianko, Andrei and Ye, Jun},
  journal = {Phys. Rev. Lett.},
  volume = {125},
  issue = {20},
  pages = {201302},
  numpages = {6},
  year = {2020},
  month = {Nov},
  publisher = {American Physical Society},
  doi = {10.1103/PhysRevLett.125.201302},
  url = {https://link.aps.org/doi/10.1103/PhysRevLett.125.201302}
}

@article{Banerjee2020,
   title={Relaxion stars and their detection via atomic physics},
   volume={3},
    pages = {1},
   ISSN={2399-3650},
   url={http://dx.doi.org/10.1038/s42005-019-0260-3},
   DOI={10.1038/s42005-019-0260-3},
   number={1},
   journal={Communications Physics},
   publisher={Springer Science and Business Media LLC},
   author={Banerjee, Abhishek and Budker, Dmitry and Eby, Joshua and Kim, Hyungjin and Perez, Gilad},
   year={2020},
   month=jan }

@article{YbCs,
  title = {Search for ultralight dark matter from long-term frequency comparisons of optical and microwave atomic clocks},
  author = {Kobayashi, Takumi and Takamizawa, Akifumi and Akamatsu, Daisuke and Kawasaki, Akio and Nishiyama, Akiko and Hosaka, Kazumoto and Hisai, Yusuke and Wada, Masato and Inaba, Hajime and Tanabe, Takehiko and Yasuda, Masami},
  journal = {Phys. Rev. Lett.},
  volume = {129},
  issue = {24},
  pages = {241301},
  numpages = {7},
  year = {2022},
  month = {Dec},
  publisher = {American Physical Society},
  doi = {10.1103/PhysRevLett.129.241301},
  url = {https://link.aps.org/doi/10.1103/PhysRevLett.129.241301}
}

@article{Thomas,
    author = "Flambaum, V. V. and Leinweber, Derek Bruce and Thomas, Anthony William and Young, Ross Daniel",
    title = "{Limits on the temporal variation of the fine structure constant, quark masses and strong interaction from quasar absorption spectra and atomic clock experiments}",
    doi = "10.1103/PhysRevD.69.115006",
    journal = "Phys. Rev. D",
    volume = "69",
    pages = "115006",
    year = "2004"
}

@article{DyCs,
  title = {Search for ultralight scalar dark matter with atomic spectroscopy},
  author = {Van Tilburg, Ken and Leefer, Nathan and Bougas, Lykourgos and Budker, Dmitry},
  journal = {Phys. Rev. Lett.},
  volume = {115},
  issue = {1},
  pages = {011802},
  numpages = {5},
  year = {2015},
  month = {Jun},
  publisher = {American Physical Society},
  doi = {10.1103/PhysRevLett.115.011802},
  url = {https://link.aps.org/doi/10.1103/PhysRevLett.115.011802}
}

@article{Antypas1,
author = {Antypas, Dionysios and Budker, Dmitry and Flambaum, Victor V. and Kozlov, Mikhail G. and Perez, Gilad and Ye, Jun},
title = {Fast Apparent Oscillations of Fundamental Constants},
journal = {Annalen der Physik},
volume = {532},
number = {4},
pages = {1900566},
doi = {https://doi.org/10.1002/andp.201900566},
url = {https://onlinelibrary.wiley.com/doi/abs/10.1002/andp.201900566},
year = {2020}
}

@article{Antypas2,
doi = {10.1088/2058-9565/abe472},
url = {https://dx.doi.org/10.1088/2058-9565/abe472},
year = {2021},
month = {apr},
publisher = {IOP Publishing},
volume = {6},
number = {3},
pages = {034001},
author = {Antypas, Dionysios and Tretiak, Oleg and Zhang, Ke and Garcon, Antoine and Perez, Gilad and Kozlov, Mikhail G. and Schiller, Stephan and Budker, Dmitry},
title = {Probing fast oscillating scalar dark matter with atoms and molecules},
journal = {Quantum Science and Technology}
}

@article{b,
title = {Remarks on Higgs-boson interactions with nucleons},
journal = {Phys. Lett. B},
volume = {78},
number = {4},
pages = {443-446},
year = {1978},
issn = {0370-2693},
doi = {https://doi.org/10.1016/0370-2693(78)90481-1},
url = {https://www.sciencedirect.com/science/article/pii/0370269378904811},
author = {M.A. Shifman and A.I. Vainshtein and V.I. Zakharov}
}

@article{Litvinova,
  title = {Nuclear structure of lowest $^{229}\mathrm{Th}$ states and time-dependent fundamental constants},
  author = {Litvinova, Elena and Feldmeier, Hans and Dobaczewski, Jacek and Flambaum, Victor},
  journal = {Phys. Rev. C},
  volume = {79},
  issue = {6},
  pages = {064303},
  numpages = {12},
  year = {2009},
  month = {Jun},
  publisher = {American Physical Society},
  doi = {10.1103/PhysRevC.79.064303},
  url = {https://link.aps.org/doi/10.1103/PhysRevC.79.064303}
}

@article{Tretiak,
  title = {Improved Bounds on Ultralight Scalar Dark Matter in the Radio-Frequency Range},
  author = {Tretiak, Oleg and Zhang, Xue and Figueroa, Nataniel L. and Antypas, Dionysios and Brogna, Andrea and Banerjee, Abhishek and Perez, Gilad and Budker, Dmitry},
  journal = {Phys. Rev. Lett.},
  volume = {129},
  issue = {3},
  pages = {031301},
  numpages = {7},
  year = {2022},
  month = {Jul},
  publisher = {American Physical Society},
  doi = {10.1103/PhysRevLett.129.031301},
  url = {https://link.aps.org/doi/10.1103/PhysRevLett.129.031301}
}

@Inbook{Fischer2004,
author="Fischer, M.
and Kolachevsky, N.
and Zimmermann, M.
and Holzwarth, R.
and Udem, Th.
and H{\"a}nsch, T.W.
and Abgrall, M.
and Gr{\"u}nert, J.
and Maksimovic, I.
and Bize, S.
and Marion, H.
and Pereira Dos Santos, F.
and Lemonde, P.
and Santarelli, G.
and Laurent, P.
and Clairon, A.
and Salomon, C.",
editor="Karshenboim, Savely G.
and Peik, Ekkehard",
title="Precision Spectroscopy of Atomic Hydrogen and Variations of Fundamental Constants",
bookTitle="Astrophysics, Clocks and Fundamental Constants",
year="2004",
publisher="Springer Berlin Heidelberg",
address="Berlin, Heidelberg",
pages="209--227",
isbn="978-3-540-40991-5",
doi="10.1007/978-3-540-40991-5_13",
url="https://doi.org/10.1007/978-3-540-40991-5_13"
}

@article{
AlHgdrift,
author = {T. Rosenband  and D. B. Hume  and P. O. Schmidt  and C. W. Chou  and A. Brusch  and L. Lorini  and W. H. Oskay  and R. E. Drullinger  and T. M. Fortier  and J. E. Stalnaker  and S. A. Diddams  and W. C. Swann  and N. R. Newbury  and W. M. Itano  and D. J. Wineland  and J. C. Bergquist },
title = {Frequency Ratio of {Al}$^{+}$ and {Hg}$^{+}$ Single-Ion Optical Clocks; Metrology at the 17th Decimal Place},
journal = {Science},
volume = {319},
number = {5871},
pages = {1808-1812},
year = {2008},
doi = {10.1126/science.1154622},
URL = {https://www.science.org/doi/abs/10.1126/science.1154622},
eprint = {https://www.science.org/doi/pdf/10.1126/science.1154622}
}

@article{Wineland2007,
  title = {Observation of the $^{1}S_{0}\ensuremath{\rightarrow}^{3}P_{0}$ Clock Transition in $^{27}\mathrm{Al}^{+}$},
  author = {Rosenband, T. and Schmidt, P. O. and Hume, D. B. and Itano, W. M. and Fortier, T. M. and Stalnaker, J. E. and Kim, K. and Diddams, S. A. and Koelemeij, J. C. J. and Bergquist, J. C. and Wineland, D. J.},
  journal = {Phys. Rev. Lett.},
  volume = {98},
  issue = {22},
  pages = {220801},
  numpages = {4},
  year = {2007},
  month = {May},
  publisher = {American Physical Society},
  doi = {10.1103/PhysRevLett.98.220801},
  url = {https://link.aps.org/doi/10.1103/PhysRevLett.98.220801}
}

@article{Tang2021,
doi = {10.1088/1674-1056/ac0130},
url = {https://dx.doi.org/10.1088/1674-1056/ac0130},
year = {2021},
month = {dec},
publisher = {Chinese Physical Society and IOP Publishing Ltd},
volume = {30},
number = {12},
pages = {123204},
author = {Xiao-Kang Tang and Xiang Zhang and Yong Shen and Hong-Xin Zou},
title = {Theoretical calculations of hyperfine splitting, Zeeman shifts, and isotope shifts of 27Al+ and logical ions in Al+ clocks*},
journal = {Chinese Physics B}
}

@article{FlambaumMansour2023,
  title = {Variation of the Quadrupole Hyperfine Structure and Nuclear Radius due to an Interaction with Scalar and Axion Dark Matter},
  author = {Flambaum, V. V. and Mansour, A. J.},
  journal = {Phys. Rev. Lett.},
  volume = {131},
  issue = {11},
  pages = {113004},
  numpages = {5},
  year = {2023},
  month = {Sep},
  publisher = {American Physical Society},
  doi = {10.1103/PhysRevLett.131.113004},
  url = {https://link.aps.org/doi/10.1103/PhysRevLett.131.113004}
}

@article{Lange2021,
  title = {Improved Limits for Violations of Local Position Invariance from Atomic Clock Comparisons},
  author = {Lange, R. and Huntemann, N. and Rahm, J. M. and Sanner, C. and Shao, H. and Lipphardt, B. and Tamm, Chr. and Weyers, S. and Peik, E.},
  journal = {Phys. Rev. Lett.},
  volume = {126},
  issue = {1},
  pages = {011102},
  numpages = {6},
  year = {2021},
  month = {Jan},
  publisher = {American Physical Society},
  doi = {10.1103/PhysRevLett.126.011102},
  url = {https://link.aps.org/doi/10.1103/PhysRevLett.126.011102}
}

@article{DZUBAPRA1999,
  title = {Calculations of the relativistic effects in many-electron atoms and space-time variation of fundamental constants},
  author = {Dzuba, V. A. and Flambaum, V. V. and Webb, J. K.},
  journal = {Phys. Rev. A},
  volume = {59},
  issue = {1},
  pages = {230--237},
  numpages = {0},
  year = {1999},
  month = {Jan},
  publisher = {American Physical Society},
  doi = {10.1103/PhysRevA.59.230},
  url = {https://link.aps.org/doi/10.1103/PhysRevA.59.230}
}

@article{DZUBAPRL1999,
  title = {Space-Time Variation of Physical Constants and Relativistic Corrections in Atoms},
  author = {Dzuba, V. A. and Flambaum, V. V. and Webb, J. K.},
  journal = {Phys. Rev. Lett.},
  volume = {82},
  issue = {5},
  pages = {888--891},
  numpages = {0},
  year = {1999},
  month = {Feb},
  publisher = {American Physical Society},
  doi = {10.1103/PhysRevLett.82.888},
  url = {https://link.aps.org/doi/10.1103/PhysRevLett.82.888}
}

@article{Uzan,
  title = {Varying Constants, Gravitation and Cosmology},
  author = {Uzan, J. P.},
  journal = {Living Reviews in Relativity},
  volume = {14},
  pages = {2},
  year = {2011},
  doi = {https://doi.org/10.12942/lrr-2011-2},
  url = {https://link.springer.com/article/10.12942/lrr-2011-2}
}

@article{Junnarkar2013,
  title = {Scalar strange content of the nucleon from lattice QCD},
  author = {Junnarkar, P. M. and Walker-Loud, A.},
  journal = {Phys. Rev. D},
  volume = {87},
  issue = {11},
  pages = {114510},
  numpages = {15},
  year = {2013},
  month = {Jun},
  publisher = {American Physical Society},
  doi = {10.1103/PhysRevD.87.114510},
  url = {https://link.aps.org/doi/10.1103/PhysRevD.87.114510}
}

@article{GellMan1968,
  title = {Behavior of Current Divergences under ${\mathrm{SU}}_{3}\ifmmode\times\else\texttimes\fi{}{\mathrm{SU}}_{3}$},
  author = {Gell-Mann, Murray and Oakes, R. J. and Renner, B.},
  journal = {Phys. Rev.},
  volume = {175},
  issue = {5},
  pages = {2195--2199},
  numpages = {0},
  year = {1968},
  month = {Nov},
  publisher = {American Physical Society},
  doi = {10.1103/PhysRev.175.2195},
  url = {https://link.aps.org/doi/10.1103/PhysRev.175.2195}
}

@article{DzubaFlambaumMansour2024,
  title = {Constraints on the variation of physical constants, equivalence principle violation, and a fifth force from atomic experiments},
  author = {Dzuba, V. A. and Flambaum, V. V. and Mansour, A. J.},
  journal = {Phys. Rev. D},
  volume = {110},
  issue = {5},
  pages = {055022},
  numpages = {13},
  year = {2024},
  month = {Sep},
  publisher = {American Physical Society},
  doi = {10.1103/PhysRevD.110.055022},
  url = {https://link.aps.org/doi/10.1103/PhysRevD.110.055022}
}

@article{Fuchs2024PRX,
  title = {Searching for Dark Matter with the $^{229}\mathrm{Th}$ Nuclear Lineshape from Laser Spectroscopy},
  author = {Fuchs, Elina and Kirk, Fiona and Madge, Eric and Paranjape, Chaitanya and Peik, Ekkehard and Perez, Gilad and Ratzinger, Wolfram and Tiedau, Johannes},
  journal = {Phys. Rev. X},
  volume = {15},
  issue = {2},
  pages = {021055},
  numpages = {15},
  year = {2025},
  month = {May},
  publisher = {American Physical Society},
  doi = {10.1103/PhysRevX.15.021055},
  url = {https://link.aps.org/doi/10.1103/PhysRevX.15.021055}
}

@article{Shuryak2002,
  title = {Limits on cosmological variation of strong interaction and quark masses from big bang nucleosynthesis, cosmic, laboratory and Oklo data},
  author = {Flambaum, V. V. and Shuryak, E. V.},
  journal = {Phys. Rev. D},
  volume = {65},
  issue = {10},
  pages = {103503},
  numpages = {11},
  year = {2002},
  month = {Apr},
  publisher = {American Physical Society},
  doi = {10.1103/PhysRevD.65.103503},
  url = {https://link.aps.org/doi/10.1103/PhysRevD.65.103503}
}

@article{Dmitriev2003,
  title = {Limits on cosmological variation of quark masses and strong interaction},
  author = {Dmitriev, V. F. and Flambaum, V. V.},
  journal = {Phys. Rev. D},
  volume = {67},
  issue = {6},
  pages = {063513},
  numpages = {5},
  year = {2003},
  month = {Mar},
  publisher = {American Physical Society},
  doi = {10.1103/PhysRevD.67.063513},
  url = {https://link.aps.org/doi/10.1103/PhysRevD.67.063513}
}

@article{Murphy2001,
    author = {Murphy, M.T. and Webb, J.K. and Flambaum, V.V. and Drinkwater, M.J. and Combes, F. and Wiklind, T.},
    title = {Improved constraints on possible variation of physical constants from H i 21-cm and molecular QSO absorption lines},
    journal = {Monthly Notices of the Royal Astronomical Society},
    volume = {327},
    number = {4},
    pages = {1244-1248},
    year = {2001},
    month = {11},
    issn = {0035-8711},
    doi = {10.1046/j.1365-8711.2001.04843.x},
    url = {https://doi.org/10.1046/j.1365-8711.2001.04843.x},
    eprint = {https://academic.oup.com/mnras/article-pdf/327/4/1244/3274942/327-4-1244.pdf},
}

@article{Shlyakhter1976,
    title = {Direct test of the constancy of fundamental nuclear constants},
    author = {Shlyakhter, A. I.},
    journal = {Nature},
    volume = {264},
    pages = {340},
    year = {1976},
    url = {https://www.nature.com/articles/264340a0}
}

@article{DAMOUR1996,
title = {The Oklo bound on the time variation of the fine-structure constant revisited},
journal = {Nuclear Physics B},
volume = {480},
number = {1},
pages = {37-54},
year = {1996},
issn = {0550-3213},
doi = {https://doi.org/10.1016/S0550-3213(96)00467-1},
url = {https://www.sciencedirect.com/science/article/pii/S0550321396004671},
author = {Thibault Damour and Freeman Dyson}
}

@article{FUJII2000,
title = {The nuclear interaction at Oklo 2 billion years ago},
journal = {Nuclear Physics B},
volume = {573},
number = {1},
pages = {377-401},
year = {2000},
issn = {0550-3213},
doi = {https://doi.org/10.1016/S0550-3213(00)00038-9},
url = {https://www.sciencedirect.com/science/article/pii/S0550321300000389},
author = {Yasunori Fujii and Akira Iwamoto and Tokio Fukahori and Toshihiko Ohnuki and Masayuki Nakagawa and Hiroshi Hidaka and Yasuji Oura and Peter Möller}
}

@article{BERENGUT2010114,
title = {Effect of quark mass variation on big bang nucleosynthesis},
journal = {Physics Letters B},
volume = {683},
number = {2},
pages = {114-118},
year = {2010},
issn = {0370-2693},
doi = {https://doi.org/10.1016/j.physletb.2009.12.002},
url = {https://www.sciencedirect.com/science/article/pii/S0370269309014245},
author = {J.C. Berengut and V.V. Flambaum and V.F. Dmitriev}
}

@article{Peik2003,
doi = {10.1209/epl/i2003-00210-x},
url = {https://dx.doi.org/10.1209/epl/i2003-00210-x},
year = {2003},
month = {jan},
publisher = {},
volume = {61},
number = {2},
pages = {181},
author = {E. Peik and Chr. Tamm},
title = {Nuclear laser spectroscopy  of the 3.5 eV
transition in Th-229},
journal = {Europhysics Letters}
}

@article{Zhang2024,
   title={Frequency ratio of the 229mTh nuclear isomeric transition and the 87Sr atomic clock},
   volume={633},
   ISSN={1476-4687},
   url={http://dx.doi.org/10.1038/s41586-024-07839-6},
   DOI={10.1038/s41586-024-07839-6},
   number={8028},
   journal={Nature},
   publisher={Springer Science and Business Media LLC},
   author={Zhang, Chuankun and Ooi, Tian and Higgins, Jacob S. and Doyle, Jack F. and von der Wense, Lars and Beeks, Kjeld and Leitner, Adrian and Kazakov, Georgy A. and Li, Peng and Thirolf, Peter G. and Schumm, Thorsten and Ye, Jun},
   year={2024},
   month=sep, pages={63–70} }

@article{Seiferle2019,
   title={Energy of the 229Th nuclear clock transition},
   volume={573},
   ISSN={1476-4687},
   url={http://dx.doi.org/10.1038/s41586-019-1533-4},
   DOI={10.1038/s41586-019-1533-4},
   number={7773},
   journal={Nature},
   publisher={Springer Science and Business Media LLC},
   author={Seiferle, Benedict and von der Wense, Lars and Bilous, Pavlo V. and Amersdorffer, Ines and Lemell, Christoph and Libisch, Florian and Stellmer, Simon and Schumm, Thorsten and Düllmann, Christoph E. and Pálffy, Adriana and Thirolf, Peter G.},
   year={2019},
   month=sep, pages={243–246} }

@article{Masuda2019,
   title={X-ray pumping of the 229Th nuclear clock isomer},
   volume={573},
   ISSN={1476-4687},
   url={http://dx.doi.org/10.1038/s41586-019-1542-3},
   DOI={10.1038/s41586-019-1542-3},
   number={7773},
   journal={Nature},
   publisher={Springer Science and Business Media LLC},
   author={Masuda, Takahiko and Yoshimi, Akihiro and Fujieda, Akira and Fujimoto, Hiroyuki and Haba, Hiromitsu and Hara, Hideaki and Hiraki, Takahiro and Kaino, Hiroyuki and Kasamatsu, Yoshitaka and Kitao, Shinji and Konashi, Kenji and Miyamoto, Yuki and Okai, Koichi and Okubo, Sho and Sasao, Noboru and Seto, Makoto and Schumm, Thorsten and Shigekawa, Yudai and Suzuki, Kenta and Stellmer, Simon and Tamasaku, Kenji and Uetake, Satoshi and Watanabe, Makoto and Watanabe, Tsukasa and Yasuda, Yuki and Yamaguchi, Atsushi and Yoda, Yoshitaka and Yokokita, Takuya and Yoshimura, Motohiko and Yoshimura, Koji},
   year={2019},
   month=sep, pages={238–242} }

@article{Yamaguchi2019,
  title = {Energy of the $^{229}\mathrm{Th}$ Nuclear Clock Isomer Determined by Absolute $\ensuremath{\gamma}$-ray Energy Difference},
  author = {Yamaguchi, A. and Muramatsu, H. and Hayashi, T. and Yuasa, N. and Nakamura, K. and Takimoto, M. and Haba, H. and Konashi, K. and Watanabe, M. and Kikunaga, H. and Maehata, K. and Yamasaki, N. Y. and Mitsuda, K.},
  journal = {Phys. Rev. Lett.},
  volume = {123},
  issue = {22},
  pages = {222501},
  numpages = {6},
  year = {2019},
  month = {Nov},
  publisher = {American Physical Society},
  doi = {10.1103/PhysRevLett.123.222501},
  url = {https://link.aps.org/doi/10.1103/PhysRevLett.123.222501}
}

@article{Sikorski2020,
  title = {Measurement of the $^{229}\mathrm{Th}$ Isomer Energy with a Magnetic Microcalorimeter},
  author = {Sikorsky, Tomas and Geist, Jeschua and Hengstler, Daniel and Kempf, Sebastian and Gastaldo, Loredana and Enss, Christian and Mokry, Christoph and Runke, J\"org and D\"ullmann, Christoph E. and Wobrauschek, Peter and Beeks, Kjeld and Rosecker, Veronika and Sterba, Johannes H. and Kazakov, Georgy and Schumm, Thorsten and Fleischmann, Andreas},
  journal = {Phys. Rev. Lett.},
  volume = {125},
  issue = {14},
  pages = {142503},
  numpages = {6},
  year = {2020},
  month = {Sep},
  publisher = {American Physical Society},
  doi = {10.1103/PhysRevLett.125.142503},
  url = {https://link.aps.org/doi/10.1103/PhysRevLett.125.142503}
}

@article{Minkov2021,
	author = {{Minkov, Nikolay} and {Pálffy, Adriana}},
	title = {Shape and electromagnetic properties of the 229mTh isomer},
	DOI= "10.1051/epjconf/202125202003",
	url= "https://doi.org/10.1051/epjconf/202125202003",
	journal = {EPJ Web Conf.},
	year = 2021,
	volume = 252,
	pages = "02003",
}

@article{Fadeev2020,
  title = {Sensitivity of $^{229}\mathrm{Th}$ nuclear clock transition to variation of the fine-structure constant},
  author = {Fadeev, Pavel and Berengut, Julian C. and Flambaum, Victor V.},
  journal = {Phys. Rev. A},
  volume = {102},
  issue = {5},
  pages = {052833},
  numpages = {6},
  year = {2020},
  month = {Nov},
  publisher = {American Physical Society},
  doi = {10.1103/PhysRevA.102.052833},
  url = {https://link.aps.org/doi/10.1103/PhysRevA.102.052833}
}

@misc{SafronovaYe,
      title={Fine-structure constant sensitivity of the Th-229 nuclear clock transition}, 
      author={Kjeld Beeks and Georgy A. Kazakov and Fabian Schaden and Ira Morawetz and Luca Toscani de Col and Thomas Riebner and Michael Bartokos and Tomas Sikorsky and Thorsten Schumm and Chuankun Zhang and Tian Ooi and Jacob S. Higgins and Jack F. Doyle and Jun Ye and Marianna S. Safronova},
      year={2024},
      eprint={2407.17300},
      archivePrefix={arXiv},
      primaryClass={nucl-th},
      url={https://arxiv.org/abs/2407.17300}, 
}

@article{FischerHCs2004,
  title = {New Limits on the Drift of Fundamental Constants from Laboratory Measurements},
  author = {Fischer, M. and Kolachevsky, N. and Zimmermann, M. and Holzwarth, R. and Udem, Th. and H\"ansch, T. W. and Abgrall, M. and Gr\"unert, J. and Maksimovic, I. and Bize, S. and Marion, H. and Santos, F. Pereira Dos and Lemonde, P. and Santarelli, G. and Laurent, P. and Clairon, A. and Salomon, C. and Haas, M. and Jentschura, U. D. and Keitel, C. H.},
  journal = {Phys. Rev. Lett.},
  volume = {92},
  issue = {23},
  pages = {230802},
  numpages = {4},
  year = {2004},
  month = {Jun},
  publisher = {American Physical Society},
  doi = {10.1103/PhysRevLett.92.230802},
  url = {https://link.aps.org/doi/10.1103/PhysRevLett.92.230802}
}

@article{Fortier2007,
  title = {Precision Atomic Spectroscopy for Improved Limits on Variation of the Fine Structure Constant and Local Position Invariance},
  author = {Fortier, T. M. and Ashby, N. and Bergquist, J. C. and Delaney, M. J. and Diddams, S. A. and Heavner, T. P. and Hollberg, L. and Itano, W. M. and Jefferts, S. R. and Kim, K. and Levi, F. and Lorini, L. and Oskay, W. H. and Parker, T. E. and Shirley, J. and Stalnaker, J. E.},
  journal = {Phys. Rev. Lett.},
  volume = {98},
  issue = {7},
  pages = {070801},
  numpages = {4},
  year = {2007},
  month = {Feb},
  publisher = {American Physical Society},
  doi = {10.1103/PhysRevLett.98.070801},
  url = {https://link.aps.org/doi/10.1103/PhysRevLett.98.070801}
}

@article{Blatt2008,
  title = {New Limits on Coupling of Fundamental Constants to Gravity Using $^{87}\mathrm{Sr}$ Optical Lattice Clocks},
  author = {Blatt, S. and Ludlow, A. D. and Campbell, G. K. and Thomsen, J. W. and Zelevinsky, T. and Boyd, M. M. and Ye, J. and Baillard, X. and Fouch\'e, M. and Le Targat, R. and Brusch, A. and Lemonde, P. and Takamoto, M. and Hong, F.-L. and Katori, H. and Flambaum, V. V.},
  journal = {Phys. Rev. Lett.},
  volume = {100},
  issue = {14},
  pages = {140801},
  numpages = {4},
  year = {2008},
  month = {Apr},
  publisher = {American Physical Society},
  doi = {10.1103/PhysRevLett.100.140801},
  url = {https://link.aps.org/doi/10.1103/PhysRevLett.100.140801}
}

@article{Gould2006,
  title = {Time variability of \ensuremath{\alpha} from realistic models of Oklo reactors},
  author = {Gould, C. R. and Sharapov, E. I. and Lamoreaux, S. K.},
  journal = {Phys. Rev. C},
  volume = {74},
  issue = {2},
  pages = {024607},
  numpages = {10},
  year = {2006},
  month = {Aug},
  publisher = {American Physical Society},
  doi = {10.1103/PhysRevC.74.024607},
  url = {https://link.aps.org/doi/10.1103/PhysRevC.74.024607}
}

@article{Petrov2006,
  title = {Natural nuclear reactor at Oklo and variation of fundamental constants: Computation of neutronics of a fresh core},
  author = {Petrov, Yu. V. and Nazarov, A. I. and Onegin, M. S. and Petrov, V. Yu. and Sakhnovsky, E. G.},
  journal = {Phys. Rev. C},
  volume = {74},
  issue = {6},
  pages = {064610},
  numpages = {17},
  year = {2006},
  month = {Dec},
  publisher = {American Physical Society},
  doi = {10.1103/PhysRevC.74.064610},
  url = {https://link.aps.org/doi/10.1103/PhysRevC.74.064610}
}

@article{Bernard1992,
  author = {Bernard, V. and Kaiser, N. and Meißner, Ulf-G.},
  title = {Chiral corrections to the nucleon magnetic moments},
  journal = {Physical Review Letters},
  volume = {69},
  number = {12},
  pages = {1877--1880},
  year = {1992},
  doi = {10.1103/PhysRevLett.69.1877}
}

@article{Bernard1995,
  author = {Bernard, V. and Kaiser, N. and Meißner, Ulf-G.},
  title = {Chiral dynamics in nucleons and nuclei},
  journal = {International Journal of Modern Physics E},
  volume = {4},
  number = {1},
  pages = {193--346},
  year = {1995},
  doi = {10.1142/S0218301395000092},
  eprint = {hep-ph/9501384}
}

@article{Kubis2001,
  author = {Kubis, Bastian and Meißner, Ulf-G.},
  title = {Low energy analysis of the nucleon electromagnetic form factors},
  journal = {Nuclear Physics A},
  volume = {679},
  pages = {698--734},
  year = {2001},
  doi = {10.1016/S0375-9474(00)00378-X},
  eprint = {hep-ph/0007056}
}

@article{Jenkins1993,
title = {Chiral perturbation theory analysis of the baryon magnetic moments},
journal = {Physics Letters B},
volume = {302},
number = {4},
pages = {482-490},
year = {1993},
issn = {0370-2693},
doi = {https://doi.org/10.1016/0370-2693(93)90430-P},
url = {https://www.sciencedirect.com/science/article/pii/037026939390430P},
author = {Elizabeth Jenkins and Michael Luke and Aneesh V. Manohar and Martin J. Savage}
}

@article{Tiedau2024,
  title = {Laser Excitation of the Th-229 Nucleus},
  author = {Tiedau, J. and Okhapkin, M. V. and Zhang, K. and Thielking, J. and Zitzer, G. and Peik, E. and Schaden, F. and Pronebner, T. and Morawetz, I. and De Col, L. Toscani and Schneider, F. and Leitner, A. and Pressler, M. and Kazakov, G. A. and Beeks, K. and Sikorsky, T. and Schumm, T.},
  journal = {Phys. Rev. Lett.},
  volume = {132},
  issue = {18},
  pages = {182501},
  numpages = {6},
  year = {2024},
  month = {Apr},
  publisher = {American Physical Society},
  doi = {10.1103/PhysRevLett.132.182501},
  url = {https://link.aps.org/doi/10.1103/PhysRevLett.132.182501}
}

@article{Katori2024,
title = "Laser spectroscopy of triply charged 229Th isomer for a nuclear clock",
author = "Atsushi Yamaguchi and Yudai Shigekawa and Hiromitsu Haba and Hidetoshi Kikunaga and Kenji Shirasaki and Michiharu Wada and Hidetoshi Katori",
note = "Publisher Copyright: {\textcopyright} The Author(s), under exclusive licence to Springer Nature Limited 2024.",
year = "2024",
month = may,
day = "2",
doi = "10.1038/s41586-024-07296-1",
language = "English",
volume = "629",
pages = "62--66",
journal = "Nature",
issn = "0028-0836",
publisher = "Nature Research",
number = "8010",
}

@article{Elwell2024,
  title = {Laser Excitation of the $^{229}\mathrm{Th}$ Nuclear Isomeric Transition in a Solid-State Host},
  author = {Elwell, R. and Schneider, Christian and Jeet, Justin and Terhune, J. E. S. and Morgan, H. W. T. and Alexandrova, A. N. and Tran Tan, H. B. and Derevianko, Andrei and Hudson, Eric R.},
  journal = {Phys. Rev. Lett.},
  volume = {133},
  issue = {1},
  pages = {013201},
  numpages = {6},
  year = {2024},
  month = {Jul},
  publisher = {American Physical Society},
  doi = {10.1103/PhysRevLett.133.013201},
  url = {https://link.aps.org/doi/10.1103/PhysRevLett.133.013201}
}

@article{Hiraki2024,
   title={Controlling 229Th isomeric state population in a VUV transparent crystal},
   volume={15},
   ISSN={2041-1723},
   url={http://dx.doi.org/10.1038/s41467-024-49631-0},
   DOI={10.1038/s41467-024-49631-0},
   number={1},
   journal={Nature Communications},
   publisher={Springer Science and Business Media LLC},
   author={Hiraki, Takahiro and Okai, Koichi and Bartokos, Michael and Beeks, Kjeld and Fujimoto, Hiroyuki and Fukunaga, Yuta and Haba, Hiromitsu and Kasamatsu, Yoshitaka and Kitao, Shinji and Leitner, Adrian and Masuda, Takahiko and Guan, Ming and Nagasawa, Nobumoto and Ogake, Ryoichiro and Pimon, Martin and Pressler, Martin and Sasao, Noboru and Schaden, Fabian and Schumm, Thorsten and Seto, Makoto and Shigekawa, Yudai and Shimizu, Kotaro and Sikorsky, Tomas and Tamasaku, Kenji and Takatori, Sayuri and Watanabe, Tsukasa and Yamaguchi, Atsushi and Yoda, Yoshitaka and Yoshimi, Akihiro and Yoshimura, Koji},
   year={2024},
   month=jul,pages = {5536} }

@article{ZhangBao2024,
   title={229ThF4 thin films for solid-state nuclear clocks},
   volume={636},
   ISSN={1476-4687},
   url={http://dx.doi.org/10.1038/s41586-024-08256-5},
   DOI={10.1038/s41586-024-08256-5},
   number={8043},
   journal={Nature},
   publisher={Springer Science and Business Media LLC},
   author={Zhang, Chuankun and von der Wense, Lars and Doyle, Jack F. and Higgins, Jacob S. and Ooi, Tian and Friebel, Hans U. and Ye, Jun and Elwell, R. and Terhune, J. E. S. and Morgan, H. W. T. and Alexandrova, A. N. and Tran Tan, H. B. and Derevianko, Andrei and Hudson, Eric R.},
   year={2024},
   month=dec, pages={603–608} }

@article{Reinhold2006,
  title = {Indication of a Cosmological Variation of the Proton-Electron Mass Ratio Based on Laboratory Measurement and Reanalysis of ${\mathrm{H}}_{2}$ Spectra},
  author = {Reinhold, E. and Buning, R. and Hollenstein, U. and Ivanchik, A. and Petitjean, P. and Ubachs, W.},
  journal = {Phys. Rev. Lett.},
  volume = {96},
  issue = {15},
  pages = {151101},
  numpages = {4},
  year = {2006},
  month = {Apr},
  publisher = {American Physical Society},
  doi = {10.1103/PhysRevLett.96.151101},
  url = {https://link.aps.org/doi/10.1103/PhysRevLett.96.151101}
}

@article{King2008,
  title = {Stringent Null Constraint on Cosmological Evolution of the Proton-to-Electron Mass Ratio},
  author = {King, Julian A. and Webb, John K. and Murphy, Michael T. and Carswell, Robert F.},
  journal = {Phys. Rev. Lett.},
  volume = {101},
  issue = {25},
  pages = {251304},
  numpages = {4},
  year = {2008},
  month = {Dec},
  publisher = {American Physical Society},
  doi = {10.1103/PhysRevLett.101.251304},
  url = {https://link.aps.org/doi/10.1103/PhysRevLett.101.251304}
}

@article{Thompson2009,
doi = {10.1088/0004-637X/703/2/1648},
url = {https://dx.doi.org/10.1088/0004-637X/703/2/1648},
year = {2009},
month = {sep},
publisher = {The American Astronomical Society},
volume = {703},
number = {2},
pages = {1648},
author = {Thompson, Rodger I. and Bechtold, Jill and Black, John H. and Eisenstein, Daniel and Fan, Xiaohui and Kennicutt, Robert C. and Martins, Carlos and Prochaska, J. Xavier and Shirley, Yancey L.},
title = {AN OBSERVATIONAL DETERMINATION OF THE PROTON TO ELECTRON MASS RATIO IN THE EARLY UNIVERSE*},
journal = {The Astrophysical Journal}
}

@article{Murphy2008,
   title={Strong Limit on a Variable Proton-to-Electron Mass Ratio from Molecules in the Distant Universe},
   volume={320},
   ISSN={1095-9203},
   url={http://dx.doi.org/10.1126/science.1156352},
   DOI={10.1126/science.1156352},
   number={5883},
   journal={Science},
   publisher={American Association for the Advancement of Science (AAAS)},
   author={Murphy, Michael T. and Flambaum, Victor V. and Muller, Sébastien and Henkel, Christian},
   year={2008},
   month=jun, pages={1611–1613} }

@article{Menten2008,
   title={Submillimeter water and ammonia absorption by the peculiarz≈0.89 interstellar medium in the gravitational lens of the PKS 1830-211 system},
   volume={492},
   ISSN={1432-0746},
   url={http://dx.doi.org/10.1051/0004-6361:200810041},
   DOI={10.1051/0004-6361:200810041},
   number={3},
   journal={Astronomy \& Astrophysics},
   publisher={EDP Sciences},
   author={Menten, K. M. and Güsten, R. and Leurini, S. and Thorwirth, S. and Henkel, C. and Klein, B. and Carilli, C. L. and Reid, M. J.},
   year={2008},
   month=oct, pages={725–730} }

@article{Henkel2009,
   title={The density, the cosmic microwave background,  and the proton-to-electron mass ratio
 in a cloud at redshift 0.9},
   volume={500},
   ISSN={1432-0746},
   url={http://dx.doi.org/10.1051/0004-6361/200811475},
   DOI={10.1051/0004-6361/200811475},
   number={2},
   journal={Astronomy \& Astrophysics},
   publisher={EDP Sciences},
   author={Henkel, C. and Menten, K. M. and Murphy, M. T. and Jethava, N. and Flambaum, V. V. and Braatz, J. A. and Muller, S. and Ott, J. and Mao, R. Q.},
   year={2009},
   month=apr, pages={725–734} }

@article{Petitjean2009,
   title={Constraining Fundamental Constants of Physics with Quasar Absorption Line Systems},
   volume={148},
   ISSN={1572-9672},
   url={http://dx.doi.org/10.1007/s11214-009-9520-y},
   DOI={10.1007/s11214-009-9520-y},
   number={1–4},
   journal={Space Science Reviews},
   publisher={Springer Science and Business Media LLC},
   author={Petitjean, Patrick and Srianand, Raghunathan and Chand, Hum and Ivanchik, Alexander and Noterdaeme, Pasquier and Gupta, Neeraj},
   year={2009},
   month=may, pages={289–300} }

@article{Malec2009,
   title={New limit on a varying proton-to-electron mass ratio from high-resolution optical quasar spectra},
   volume={5},
   ISSN={1743-9221},
   url={http://dx.doi.org/10.1017/S1743921310009580},
   DOI={10.1017/s1743921310009580},
   number={H15},
   journal={Proceedings of the International Astronomical Union},
   publisher={Cambridge University Press (CUP)},
   author={Malec, A. L. and Buning, R. and Murphy, M. T. and Milutinovic, N. and Ellison, S. L. and Prochaska, J. X. and Kaper, L. and Tumlinson, J. and Carswell, R. F. and Ubachs, W.},
   year={2009},
   month=nov, pages={321–321} }

@article{Kanekar2005,
   title={Constraints on Changes in Fundamental Constants from a Cosmologically Distant OH Absorber or Emitter},
   volume={95},
   ISSN={1079-7114},
   url={http://dx.doi.org/10.1103/PhysRevLett.95.261301},
   DOI={10.1103/physrevlett.95.261301},
   number={26},
   journal={Physical Review Letters},
   publisher={American Physical Society (APS)},
   author={Kanekar, N. and Carilli, C. L. and Langston, G. I. and Rocha, G. and Combes, F. and Subrahmanyan, R. and Stocke, J. T. and Menten, K. M. and Briggs, F. H. and Wiklind, T.},
   year={2005},
   month=dec, pages = {261301} }

@article{Darling2004,
   title={A Laboratory for Constraining Cosmic Evolution of the Fine‐Structure Constant: Conjugate 18 Centimeter OH Lines toward PKS 1413+135 atz= 0.24671},
   volume={612},
   ISSN={1538-4357},
   url={http://dx.doi.org/10.1086/422450},
   DOI={10.1086/422450},
   number={1},
   journal={The Astrophysical Journal},
   publisher={American Astronomical Society},
   author={Darling, Jeremy},
   year={2004},
   month=sep, pages={58–63} }

@article{Tzanavaris2007,
   title={Probing variations in fundamental constants with radio and optical quasar absorption-line observations},
   volume={374},
   ISSN={1365-2966},
   url={http://dx.doi.org/10.1111/j.1365-2966.2006.11178.x},
   DOI={10.1111/j.1365-2966.2006.11178.x},
   number={2},
   journal={Monthly Notices of the Royal Astronomical Society},
   publisher={Oxford University Press (OUP)},
   author={Tzanavaris, P. and Murphy, M. T. and Webb, J. K. and Flambaum, V. V. and Curran, S. J.},
   year={2007},
   month=jan, pages={634–646} }

@article{Srianand2010,
   title={Detection of 21-cm, H2 and deuterium absorption at z &gt; 3 along the line of sight to J1337+3152★: DLAs towards J1337+3152},
   ISSN={1365-2966},
   url={http://dx.doi.org/10.1111/j.1365-2966.2010.16574.x},
   DOI={10.1111/j.1365-2966.2010.16574.x},
   journal={Monthly Notices of the Royal Astronomical Society},
   publisher={Oxford University Press (OUP)},
   author={Srianand, R. and Gupta, N. and Petitjean, P. and Noterdaeme, P. and Ledoux, C.},
   year={2010},
   month=apr, pages={1888-1900} }

@article{Levshakov2008,
   title={A new approach for
testing variations of fundamental constants over cosmic epochs using FIR
fine-structure lines},
   volume={479},
   ISSN={1432-0746},
   url={http://dx.doi.org/10.1051/0004-6361:20079116},
   DOI={10.1051/0004-6361:20079116},
   number={3},
   journal={Astronomy \& Astrophysics},
   publisher={EDP Sciences},
   author={Levshakov, S. A. and Reimers, D. and Kozlov, M. G. and Porsev, S. G. and Molaro, P.},
   year={2008},
   month=jan, pages={719–723} }

@article{Chengalur2003,
   title={Constraining the Variation of Fundamental Constants using 18 cm OH Lines},
   volume={91},
   ISSN={1079-7114},
   url={http://dx.doi.org/10.1103/PhysRevLett.91.241302},
   DOI={10.1103/physrevlett.91.241302},
   number={24},
   journal={Physical Review Letters},
   publisher={American Physical Society (APS)},
   author={Chengalur, Jayaram N. and Kanekar, Nissim},
   year={2003},
   month=dec }

@article{Kanekar2010,
   title={PROBING FUNDAMENTAL CONSTANT EVOLUTION WITH REDSHIFTED CONJUGATE-SATELLITE OH LINES},
   volume={716},
   ISSN={2041-8213},
   url={http://dx.doi.org/10.1088/2041-8205/716/1/L23},
   DOI={10.1088/2041-8205/716/1/l23},
   number={1},
   journal={The Astrophysical Journal},
   publisher={American Astronomical Society},
   author={Kanekar, Nissim and Chengalur, Jayaram N. and Ghosh, Tapasi},
   year={2010},
   month=may, pages={L23–L26} }

@article{Cyburt2016,
   title={Big bang nucleosynthesis: Present status},
   volume={88},
   ISSN={1539-0756},
   url={http://dx.doi.org/10.1103/RevModPhys.88.015004},
   DOI={10.1103/revmodphys.88.015004},
pages = {015004},
  numpages = {5},
   number={1},
   journal={Reviews of Modern Physics},
   publisher={American Physical Society (APS)},
   author={Cyburt, Richard H. and Fields, Brian D. and Olive, Keith A. and Yeh, Tsung-Han},
   year={2016},
   month=feb }

@article{Fields2011,
   title={The Primordial Lithium Problem},
   volume={61},
   ISSN={1545-4134},
   url={http://dx.doi.org/10.1146/annurev-nucl-102010-130445},
   DOI={10.1146/annurev-nucl-102010-130445},
   number={1},
   journal={Annual Review of Nuclear and Particle Science},
   publisher={Annual Reviews},
   author={Fields, Brian D.},
   year={2011},
   month=nov, pages={47–68} }

@article{Dmitriev2004,
   title={Cosmological variation of the deuteron binding energy, strong interaction, and quark masses from big bang nucleosynthesis},
   volume={69},
   ISSN={1550-2368},
   url={http://dx.doi.org/10.1103/PhysRevD.69.063506},
   DOI={10.1103/physrevd.69.063506},
   number={6},
   journal={Physical Review D},
pages = {063506},
  numpages = {5},
   publisher={American Physical Society (APS)},
   author={Dmitriev, V. F. and Flambaum, V. V. and Webb, J. K.},
   year={2004},
   month=mar }

@article{Aoki2020,
   title={FLAG Review 2019: Flavour Lattice Averaging Group (FLAG)},
   volume={80},
   ISSN={1434-6052},
   url={http://dx.doi.org/10.1140/epjc/s10052-019-7354-7},
   DOI={10.1140/epjc/s10052-019-7354-7},
   number={2},
   journal={The European Physical Journal C},
   publisher={Springer Science and Business Media LLC},
   author={Aoki, S. and Aoki, Y. and Bečirević, D. and Blum, T. and Colangelo, G. and Collins, S. and Della Morte, M. and Dimopoulos, P. and Dürr, S. and Fukaya, H. and Golterman, M. and Gottlieb, Steven and Gupta, R. and Hashimoto, S. and Heller, U. M. and Herdoiza, G. and Horsley, R. and Jüttner, A. and Kaneko, T. and Lin, C.-J. D. and Lunghi, E. and Mawhinney, R. and Nicholson, A. and Onogi, T. and Pena, C. and Portelli, A. and Ramos, A. and Sharpe, S. R. and Simone, J. N. and Simula, S. and Sommer, R. and Van de Water, R. and Vladikas, A. and Wenger, U. and Wittig, H.},
   year={2020},
   month=feb, pages = {113} }

@article{Oswald2022,
  title = {Search for Dark-Matter-Induced Oscillations of Fundamental Constants Using Molecular Spectroscopy},
  author = {Oswald, R. and Nevsky, A. and Vogt, V. and Schiller, S. and Figueroa, N. L. and Zhang, K. and Tretiak, O. and Antypas, D. and Budker, D. and Banerjee, A. and Perez, G.},
  journal = {Phys. Rev. Lett.},
  volume = {129},
  issue = {3},
  pages = {031302},
  numpages = {7},
  year = {2022},
  month = {Jul},
  publisher = {American Physical Society},
  doi = {10.1103/PhysRevLett.129.031302},
  url = {https://link.aps.org/doi/10.1103/PhysRevLett.129.031302}
}

@article{Budker2023,
doi = {10.1088/1475-7516/2023/12/021},
url = {https://dx.doi.org/10.1088/1475-7516/2023/12/021},
year = {2023},
month = {dec},
publisher = {IOP Publishing},
volume = {2023},
number = {12},
pages = {021},
author = {Budker, Dmitry and Eby, Joshua and Gorghetto, Marco and Jiang, Minyuan and Perez, Gilad},
title = {A generic formation mechanism of ultralight dark matter solar halos},
journal = {Journal of Cosmology and Astroparticle Physics}
}

@article{Damour2010,
  title = {Equivalence principle violations and couplings of a light dilaton},
  author = {Damour, Thibault and Donoghue, John F.},
  journal = {Phys. Rev. D},
  volume = {82},
  issue = {8},
  pages = {084033},
  numpages = {20},
  year = {2010},
  month = {Oct},
  publisher = {American Physical Society},
  doi = {10.1103/PhysRevD.82.084033},
  url = {https://link.aps.org/doi/10.1103/PhysRevD.82.084033}
}

@article{SHIFMAN1978,
title = {Remarks on Higgs-boson interactions with nucleons},
journal = {Physics Letters B},
volume = {78},
number = {4},
pages = {443-446},
year = {1978},
issn = {0370-2693},
doi = {https://doi.org/10.1016/0370-2693(78)90481-1},
url = {https://www.sciencedirect.com/science/article/pii/0370269378904811},
author = {M.A. Shifman and A.I. Vainshtein and V.I. Zakharov}
}

@article{BanerjeeFlambaum2020,
   title={Searching for Earth/Solar axion halos},
   volume={2020},
   ISSN={1029-8479},
   url={http://dx.doi.org/10.1007/JHEP09(2020)004},
   DOI={10.1007/jhep09(2020)004},
   pages={4},
   journal={Journal of High Energy Physics},
   publisher={Springer Science and Business Media LLC},
   author={Banerjee, Abhishek and Budker, Dmitry and Eby, Joshua and Flambaum, Victor V. and Kim, Hyungjin and Matsedonskyi, Oleksii and Perez, Gilad},
   year={2020},
   month=sep }

@article{Barrow2005,
   title={Varying constants},
   volume={363},
   ISSN={1471-2962},
   url={http://dx.doi.org/10.1098/rsta.2005.1634},
   DOI={10.1098/rsta.2005.1634},
   number={1834},
   journal={Philosophical Transactions of the Royal Society A: Mathematical, Physical and Engineering Sciences},
   publisher={The Royal Society},
   author={Barrow, JohnD},
   year={2005},
   month=jul, pages={2139–2153} }

@article{BarrowMagueijo1998,
   title={Varying-$\alpha$ theories and solutions to the cosmological problems},
   volume={443},
   ISSN={0370-2693},
   url={http://dx.doi.org/10.1016/S0370-2693(98)01294-5},
   DOI={10.1016/s0370-2693(98)01294-5},
   number={1–4},
   journal={Physics Letters B},
   publisher={Elsevier BV},
   author={Barrow, John D. and Magueijo, João},
   year={1998},
   month=dec, pages={104–110} }

@article{BarrowSandvikMagueijo2002,
  title = {Behavior of varying-alpha cosmologies},
  author = {Barrow, John D. and Sandvik, H\aa{}vard Bunes and Magueijo, Jo\~ao},
  journal = {Phys. Rev. D},
  volume = {65},
  issue = {6},
  pages = {063504},
  numpages = {9},
  year = {2002},
  month = {Feb},
  publisher = {American Physical Society},
  doi = {10.1103/PhysRevD.65.063504},
  url = {https://link.aps.org/doi/10.1103/PhysRevD.65.063504}
}

@article{Wetterich2003,
   title={Crossover quintessence and cosmological history of fundamental “constants”},
   volume={561},
   ISSN={0370-2693},
   url={http://dx.doi.org/10.1016/S0370-2693(03)00383-6},
   DOI={10.1016/s0370-2693(03)00383-6},
   number={1–2},
   journal={Physics Letters B},
   publisher={Elsevier BV},
   author={Wetterich, C},
   year={2003},
   month=may, pages={10–16} }

@article{MurphyWebbFlambaum2003,
   title={Further evidence for a variable fine-structure constant from Keck/HIRES QSO absorption spectra},
   volume={345},
   ISSN={1365-2966},
   url={http://dx.doi.org/10.1046/j.1365-8711.2003.06970.x},
   DOI={10.1046/j.1365-8711.2003.06970.x},
   number={2},
   journal={Monthly Notices of the Royal Astronomical Society},
   publisher={Oxford University Press (OUP)},
   author={Murphy, M. T. and Webb, J. K. and Flambaum, V. V.},
   year={2003},
   month=oct, pages={609–638} }

@article{RbCsNew,
  title = {Improved Tests of Local Position Invariance Using $^{87}\mathrm{Rb}$ and $^{133}\mathrm{Cs}$ Fountains},
  author = {Gu\'ena, J. and Abgrall, M. and Rovera, D. and Rosenbusch, P. and Tobar, M. E. and Laurent, Ph. and Clairon, A. and Bize, S.},
  journal = {Phys. Rev. Lett.},
  volume = {109},
  issue = {8},
  pages = {080801},
  numpages = {5},
  year = {2012},
  month = {Aug},
  publisher = {American Physical Society},
  doi = {10.1103/PhysRevLett.109.080801},
  url = {https://link.aps.org/doi/10.1103/PhysRevLett.109.080801}
}

\end{document}